\documentclass[a4paper,11pt]{article}
\pdfoutput=1 

\usepackage{jheppub} 

\usepackage[T1]{fontenc} 
\usepackage{caption}
\usepackage{comment}
\usepackage{amsmath}
\usepackage{graphicx}
\usepackage{tikz}
\usetikzlibrary{arrows,decorations.pathmorphing}
\usepackage[compat=1.1.0]{tikz-feynman}
\usepackage{amssymb}
\usepackage{bm,commath,mathtools,wasysym,xfrac}
\usepackage{pdfpages}
\usepackage{tcolorbox}
\usepackage{braket}
\usepackage{MnSymbol}
\usepackage[english]{babel} 
\usepackage{blindtext}

\newcommand{\rootkerr}{\sqrt{\text{Kerr}}}

\newcommand{\bq}{\bar{q}}

\newcommand*{\Scale}[2][4]{\scalebox{#1}{$#2$}}%
\newcommand{\Ll}{\mathbb{L}}
\usepackage{cancel}
\usepackage{color}

\title{\boldmath Classical Observables using Exponentiated Spin factors: Electromagnetic Scattering}

\author[a,c]{Samim Akhtar,}
\author[b]{Arkajyoti Manna,}
\author[a,c]{and Akavoor Manu }

\affiliation[a]{The Institute of Mathematical Sciences \\
	IV Cross Road, C.I.T. Campus, Taramani, Chennai 600 113, India}
\affiliation[b]{Center for High Energy Physics, Indian Institute of Science,\\ C.V. Raman Avenue,
	Bangalore 560012, India}
\affiliation[c]{Homi Bhabha National Institute \\
	Training School Complex, Anushakti Nagar, Mumbai 400 094, India}

\emailAdd{samimakhtar@imsc.res.in}
\emailAdd{arkajyotim@iisc.ac.in}
\emailAdd{amanu@imsc.res.in}

\abstract{In \cite{Arkani-Hamed:2019ymq}, the authors argued that the Newman-Janis algorithm on the space of classical solutions in general relativity and electromagnetism could be used in the space of scattering amplitudes to map an amplitude with external scalar states to an amplitude associated to the scattering of ``infinite spin particles''. The minimal coupling of these particles to the gravitational or Maxwell field is equivalent to the classical coupling of the Kerr blackhole with linearized gravity or the so-called $\rootkerr$ charged state with the electromagnetic field. The action of the Newman-Janis mapping on scattering amplitudes was then used to compute the linear impulse at first post-Minkowskian (1PM) order, via the Kosower, Maybee, O'Connell (KMOC) formalism. In this paper, we continue with the idea of using the Newman-Janis mapping on the space of scalar QED amplitudes to compute classical observables such as the radiative gauge field and the angular impulse. We show that for tree-level amplitudes, the Newman-Janis action can be reinterpreted as a dressing of the photon propagator. 
	This turns out to be an efficient way to compute these  classical observables. Along the way, we highlight a  subtlety that arises in proving the conservation of angular momentum for scalar $- \rootkerr$ scattering.}

\makeatletter
\gdef\@fpheader{}
\makeatother

\begin{document} 
	\maketitle
	\flushbottom
	
	\section{Introduction}
	The array of tools available at our disposal in computing observables in gravitational and electromagnetic scattering has seen a remarkable increase in the last few years. A central premise in this resurgence is the realization that on-shell amplitudes in gauge theories and gravity can be deployed to compute classical observables ranging from scattering angle to radiative flux in Post Minkowskian scattering \cite{Bern:2019nnu,Bern:2019crd,Bern:2020gjj,Bern:2020buy,Bern:2020uwk,Bern:2021dqo,Bern:2021yeh,Bern:2021xze,Bern:2022kto,Bern:2022jvn,Barack:2023oqp,Bern:2023ccb,Manohar:2022dea,KoemansCollado:2019ggb,Cristofoli:2020uzm,DiVecchia:2021ndb,DiVecchia:2021bdo,Alessio:2022kwv,DiVecchia:2022piu,DiVecchia:2022owy,Bjerrum-Bohr:2018xdl,Cristofoli:2019neg,Bjerrum-Bohr:2019kec,Bjerrum-Bohr:2021vuf,Bjerrum-Bohr:2021din,Bjerrum-Bohr:2021wwt,Menezes:2022tcs,Kalin:2019rwq,Cho:2021arx,Kalin:2019inp,Kalin:2020fhe,Kalin:2020lmz,Dlapa:2021npj,Dlapa:2021vgp,Dlapa:2022lmu,Dlapa:2023hsl,Kalin:2020mvi,Liu:2021zxr,Kalin:2022hph}. One of the most potent formalisms which uses the quantum S-matrix to generate classical observables was proposed in a seminal paper by Kosower, Maybee, and O'Connell (KMOC) \cite{Kosower:2018adc,Cristofoli:2021vyo,Maybee:2019jus,Aoude:2021oqj,Adamo:2022rmp}. The KMOC formalism uses scattering amplitudes to compute a set of ``in-in'' asymptotic quantities, whose classical limits are the observables of interest.  
	
	In addition to a large class of state of the art techniques which have been developed in the last several decades to analyze perturbative amplitudes, there are a few non-perturbative results in the study of amplitudes (such as soft factorization theorems \cite{Sen:2017nim,Laddha:2017ygw,Chakrabarti:2017ltl,Laddha:2018rle,Laddha:2018myi,Sahoo:2018lxl,Laddha:2019yaj,Saha:2019tub,Ghosh:2021bam,Sahoo:2020ryf}) which can be used to significantly simplify the computation of the same.
	
	Among the latter category of tools, a technique was proposed in \cite{Arkani-Hamed:2019ymq} which ``spins'' the external states in scalar QED or scalar-GR amplitudes from scalars to massive infinite spin particles that are coupled minimally to photons or gravitons. These states are known as $\sqrt{\textrm{Kerr}}$ or $\textrm{Kerr}$ particles respectively. The ``spinning'' technique discovered in \cite{Arkani-Hamed:2019ymq} was inspired by the well-known Newman-Janis (NJ) algorithm in classical general relativity and electromagnetism. The NJ algorithm implements a remarkable symmetry on the space of solutions of general relativity/electromagnetism. As Newman and Janis showed, a complex coordinate transformation can be used to derive the metric of a Kerr black hole from the metric of the Schwarzschild solution \cite{nj}.  Remarkably,  one would get the Kerr solution by simply deforming the radial coordinate $r$ in the Schwarzschild solution with an imaginary factor involving the ring radius of the Kerr black hole.  Thereafter, various kinds of Newman-Janis shifts have been explored at great length \cite{Erbin:2016lzq}- providing specific ways to implement such complex deformation to derive classical solutions for different spinning objects. In their original work, Newman and Janis also noted that the same shift on the Coulombic field produces, what is now known, as the $\rootkerr$ field. The features of this solution were further expanded upon in \cite{Lynden-Bell:2002dvr}.
		
	Motivated by the ideas in \cite{Alessio:2023kgf}, we propose a ``spin dressed'' photon propagator and use this to compute the following observables in a scattering process involving a scalar and a $\rootkerr$ particle\footnote{We emphasize that we do not use the Lagrangian proposed in \cite{Alessio:2023kgf}, which leads to incorrect opposite helicity Compton amplitude for the spinning particle. We use only the exponentiation of the minimally coupled 3-point amplitudes \cite{Arkani-Hamed:2019ymq} to compute all the observables in this paper.}. All our computations are at leading order in the coupling.\footnote{This is the analog of leading order (LO) post-Minkowskian expansion in gravity.}
	\begin{enumerate}
		\item The electromagnetic radiation emitted by the scalar particle. This computation shows the power of the NJ algorithm even in the non-conservative sector. We show that just as for the linear impulse, the radiation emitted by scalar particle can also be obtained via complexification of the impact parameter $\vec{b}\, \rightarrow\, \vec{b} + i \vec{a}$.
		\item We use the NJ shift to compute the angular impulse for the scalar and $\rootkerr$ particles. We also highlight an important subtlety that underlies the computation of angular impulse for the spinning particle. This subtlety is crucially tied to the use of spin tensor ($S^{\mu \nu}$) as opposed to the spin pseudovector ($a^\mu$) as the fundamental variable. These are related via the following duality relation
		\begin{align}\label{11}
			a^\mu =\frac{1}{2m^2}\epsilon ^{\mu \nu \rho \sigma}p_{\nu}S_{\rho \sigma}\,.
		\end{align}
		\item We show that to linear order in the initial spin parameter, the total angular impulse (of the scalar-$\rootkerr$ system) is consistent with classical results \cite{Bern:2023ity} so long as the initial coherent state of the spinning particle is parametrized in terms of $S_{\mu\nu}$ and $p^{\mu}$. 
	\end{enumerate}

	The plan of the paper is as follows.  In section \ref{KMOCsetup}, we review the KMOC formalism used to compute classical observables from scattering amplitudes. In section \ref{NJ_algorithm}, we give a brief review of the Newman-Janis algorithm used as a classical solution-generating technique and its manifestation at the level of three-point amplitudes. In section \ref{sec:lin_imp}, we reformulate the NJ algorithm for the case of scattering amplitudes via a specific deformation of the photon polarisation data. We then use this version of the NJ algorithm to compute the radiation emitted by the scalar particle in section \ref{rad_ker}. In section \ref{sec7} we compute the orbital angular impulse suffered by the scalar particle, the spin angular momentum change for the $\rootkerr$ particle and highlight an important subtlety involved in computing the orbital angular momentum of the $\rootkerr$ particle. We conclude by discussing some open questions in section \ref{discussion}. In the appendices we state our conventions, provide the classical calculations for the observables and evaluate some integrals that are used in the main text.	
	
	\section{KMOC Formalism in a nut-shell} \label{KMOCsetup}
	The KMOC formalism \cite{Kosower:2018adc,Cristofoli:2021vyo} is a framework that is used to compute classical observables from on-shell scattering amplitudes for large impact parameter scattering\footnote{The scattering setups in which the particles don't deviate from the initial trajectories very much are known as large impact parameter scattering. Naturally, the characteristic length scale is set by the impact parameter. Recently, in \cite{Cristofoli:2021vyo}, the formalism has been extended beyond this regime for incoming waves scattering off massive particles.}. The procedure to compute classical observables is as follows: we start with an initial coherent state, compute the change in the expectation value of a self-adjoint quantum mechanical operator, and then take an appropriate classical limit to obtain the classical result. The main feature of the formalism is that the classical limit is taken before evaluating the full amplitude because of which the computation becomes significantly simpler. Additionally, radiation reaction effects are naturally inbuilt within the framework. For a short sample of the results obtained with the formalism, we refer the reader to \cite{Bautista:2021llr,Elkhidir:2023dco,Bern:2021xze}. In this section, we shall highlight some of the features of the formalism relevant to us.
	
	We start by describing the initial state,  
	\begin{equation}\label{ini}
		\ket{\Psi} =\int\prod_{i=1}^{2}d\Phi(p_{i}) e^{ip_{2}\cdot b/\hbar}\phi_{i}(p_{i})  \zeta^{a_{i}}_{i}\ket{\vec{p}_{1},a_{1};\vec{p}_{2},a_{2}},
	\end{equation}
	where
	\begin{equation}
		d\Phi(p) = \frac{d^{4}p}{(2\pi)^{4}}\hat{\delta}(p^{2}-m^{2})\Theta(p^{0}), \  \int d\Phi(p)\ |\phi(p)|^{2} = 1,
	\end{equation}
	$\phi_{i}(p_{i})$s are the minimum uncertainty wave packets (in momentum space) and $\zeta_{i}^{a_{i}}$ are the coherent spin state wave function for the particles with the little group indices for the particles being denoted by $a_{i}$. The wavepacket of the second particle is translated, with respect to the first particle's wavepacket, by a distance of $b$ - the impact parameter. Since the initial particles are described by coherent states we have
	\begin{equation}
		\braket{\mathbb{P}^{\mu}_{i}} = m_{i}u^{\mu}_{i}+\mathcal{O}(\hbar),\qquad \frac{\sigma_{i}^{2}}{m_{i}^{2}} = \frac{(\braket{\mathbb{P}^{2}_{i}} - \braket{\mathbb{P}_{i}}^{2})}{m_{i}^{2}} \xrightarrow{\hbar \rightarrow 0} 0,
	\end{equation}
	where $\sigma_{i}^{2}$ is the variance and $m_{i}$s are the masses of the particles. Here the expectation value of the momentum operator is with respect to the initial state in eq.\eqref{ini}.	
	
	The spin of a particle in quantum field theory is given by the expectation value of the Pauli-Lubanski vector \cite{Maybee:2019jus}, 
	\begin{align}
		\mathbb{W}^{\mu}=\frac{1}{m}\epsilon^{\mu\nu\rho\sigma}\mathbb{P}_{\nu}\mathbb{S}_{\rho\sigma} \,.
	\end{align}
	Hence, it is the expectation value of the above operator which gives the classical spin pseudovector,
	\begin{align}
		\braket{\mathbb{W}^{\mu}_{i}} = s_{i}^{\mu} \, +\mathcal{O}(\hbar).
	\end{align}    
	The variance for the spin is also small, as for the momentum of the particle. For a more detailed construction of these wavefunctions, we refer the reader to \cite{Aoude:2021oqj}.\\
	We now move on to describe the construction of the classical observables. The basic idea is to compute the change in the expectation value of a quantum mechanical operator as this is what is relevant from a classical perspective. So we write 
	\begin{align}\label{3.1}
		\braket{\Delta\mathbb{O}^{A}}=\bra{\Psi}S^{\dag}\mathbb{O}^{A}S\ket{\Psi} - \bra{\Psi}\mathbb{O}^{A}\ket{\Psi}.
	\end{align}
	where $S=I+iT$ is the S-matrix. For the linear impulse, $\mathbb{O}^{A}=\mathbb{P}^{\mu}$, the momentum operator, $\mathbb{O}^{A}=\mathbb{J}^{\mu\nu}$ for angular impulse, $\mathbb{O}^{A} = \mathbb{W}^{\mu}/m$ for the spin kick and $\mathbb{O}^{A}=\mathbb{A}^{\mu}(x)$, the gauge field operator, from which we read off the radiation kernel.
	The expression in eq.\eqref{3.1} can be simplified using the on-shell completeness relation and unitarity of the S-matrix \cite{Kosower:2018adc}. For the linear impulse, we get
	\begin{align}
		\Scale[0.95]{\langle \Delta p^{\mu} \rangle =  \int \hat{d}^{4}q\ \hat{\delta}(2p_{1}\cdot q + q^{2})\hat{\delta}(2p_{2}\cdot q-q^{2})\ e^{iq\cdot b/\hbar} q^{\mu} \mathcal{A}_{4}(p_{1},p_{2}\rightarrow p_{1}+q,p_{2}-q)  +\mathcal{O}(T^{\dag}T)}
	\end{align}
	and for the orbital angular impulse, we have 
	\begin{equation}\label{ang_momentum_exp}
		\begin{split}
			\langle \Delta L^{\mu\nu} \rangle = &  \hbar \int \prod_{i=1}^{2}\ \hat{d}^{4}q_{i}\ \hat{\delta}(2p_{i}\cdot q_{i} + q_{i}^{2})\ e^{-iq_{2}\cdot b/\hbar}\\   &\Scale[0.95]{\sum_{j=1}^{2} \bigg(\bigg(p_{j}\wedge \frac{\partial}{\partial p_{j}}\bigg)^{\mu\nu}+\bigg((p_{j}+q_{j})\wedge \frac{\partial}{\partial (p_{j}+q_{j})}\bigg)^{\mu\nu}\bigg) 
				A_{4}(p_{1},p_{2}\rightarrow p_{1}+q_{1},p_{2}+q_{2})\ \, +\mathcal{O}(T^{\dag}T)},
		\end{split}
	\end{equation}
	where we write the full 4-point amplitude as
	\begin{align}
		A_4(p_{1},p_{2}\rightarrow p_{1}+q_{1},p_{2}+q_{2}) :=\mathcal{A}_{4}(p_{1},p_{2}\rightarrow p_{1}+q_{1},p_{2}+q_{2}) \hat{\delta}^{(4)}(q_1+q_2)\,.
	\end{align}
	We provide a derivation of the above expression in Appendix \ref{angularmomentumderiv}. We do not display the higher order contributions as in this work we will only be interested in calculating the observables to leading order in coupling. Similarly, for the gauge field, we can read off the radiation kernel \cite{Cristofoli:2021vyo}
	\begin{align}
		\langle \mathcal{R}^{\mu}(k) \rangle &=\ \hbar^{3/2}\int \prod_{i=1}^{2} \hat{d}^{4}q_{i}\  \hat{\delta}(2p_{i}\cdot q_{i}+q_{i}^{2})\ e^{-iq_{2}\cdot b/\hbar}\cr
		&\qquad \qquad \hat{\delta}^{(4)}(q_{1}+q_{2}-k)\ \mathcal{A}_{5}(p_{1}+q_{1},p_{2}+q_{2}\rightarrow p_{1},p_{2,},k)\  +\ \mathcal{O}(T^{\dag}T)\,.
	\end{align}
	The spin kick, to leading order in the coupling, is given by \cite{Maybee:2019jus}
	\begin{align}
		\langle \Delta a^{\mu} \rangle &= \int \hat{d}^{4}q\ \hat{\delta}(2p_{1}\cdot q + q^{2})\hat{\delta}(2p_{2}\cdot q-q^{2})\ e^{iq\cdot b/\hbar} \cr
		&\qquad \qquad \qquad \Big(a^{\mu}(p+ q)\mathcal{A}_{4}(p_{1},p_{2}\rightarrow p_{1}+q,p_{2}-q) - \mathcal{A}_{4}(p_{1},p_{2}\rightarrow p_{1}+q,p_{2}-q)\ a^{\mu}(p)\Big) \,. \cr
	\end{align} 
	The classical limit is taken at the level of integrand, by expressing massless momenta in terms of their wave numbers (e.g.\ $q_{i}=\bar{q}_{i}\hbar$), appropriately rescaling the dimensionful couplings and keeping leading order terms in $\hbar$. In QED, the dimensionful coupling is obtained by\footnote{The dimensionless coupling is the fine structure constant $\alpha=\frac{e^{2}}{\hbar}$.} $e\rightarrow e/\sqrt{\hbar}$. 
	
	For spinning external states, in the classical limit  the final state spin pseudovector can be written as
	\begin{align}
		a_{i}^{\mu}(p_{i}+\hbar \bq_{i}) &= a_{i}^{\mu}(p_{i}) + \Delta a_{i}^{\mu}(p_{i})\,, \cr
		\Delta a_{i}^{\mu}(p_{i}) & = \omega^{\mu}_{\ \nu}(p_{i};\bar{q}_{i})a^{\nu}(p) \,, \cr
		\omega^{\mu}_{\ \nu}(p_{i};\bar{q}_{i}) &=-\frac{\hbar}{m^{2}}(p_{i}\wedge \bar{q})^{\mu}_{\ \nu} \,,
	\end{align}
	where $\omega^{\mu\nu}(p;\bar{q})$ is the infinitesimal boost parameter. With these expressions in hand, we can write down the classical limit of the linear impulse, the radiation kernel, and the spin kick, at leading order in the coupling. The expression for the former is
	\begin{align} \label{linearimpKMOCformula}
		\Delta p^{\mu}  = \Big \llangle \int \hat{d}^{4}\bar{q}\ \hat{\delta}(2p_{1}\cdot \bar{q} )\hat{\delta}(2p_{2}\cdot \bar{q})\ e^{i\bar{q}\cdot b}\ \bar{q}^{\mu}\ (\hbar^{2}\mathcal{A}_{4}(p_{1},p_{2}\rightarrow p_{1}+\hbar\bar{q},p_{2}-\hbar\bar{q})) \Big \rrangle \,.
	\end{align}
	Here $\Big \llangle f(p_{1},p_{2},q\ldots)\Big \rrangle$ denotes the integration over the minimum uncertainty wave packets which localizes the momenta and spin onto their classical values. 
	Similarly, for the radiation kernel, we get
	\begin{align} \label{radiationkernel}
		\mathcal{R}^{\mu}(\bar{k})=  \Big \llangle \int \prod_{i=1}^{2} \hat{d}^{4}\bar{q}_{i}\  \hat{\delta}(2p_{i}\cdot \bar{q}_{i}) & \ e^{-i\bar{q}_{2}\cdot b}\ \hat{\delta}^{(4)}(\bq_{1} +\bq_{2} -\bar{k})\cr
		& \hspace{1cm}\big(\hbar^{2}\mathcal{A}_{5}(p_{1}+\hbar\bar{q}_{1},p_{2}+\hbar\bar{q}_{2}\rightarrow p_{1},p_{2,},\hbar\bar{k})\big)\Big \rrangle \,,
	\end{align}
	and for the spin kick, we get
	\begin{align} 
		\Delta a^\mu &= \Big \llangle i\hbar^2 \int \hat{d}^4 \bq \hat{\delta}(2 p_1 \cdot \bq)\hat{\delta}( 2 p_2 \cdot \bq) e^{i\bq \cdot b} \Big\{ \left[a^\mu (p) , \mathcal{A}_4 \right] + \frac{\hbar}{m}(a\cdot \bq)u^\mu  \mathcal{A}_4  \Big\} \Big \rrangle \,.
	\end{align} 
	In all of the above expressions, we have taken out the $\hbar-$ scaling of the coupling constant. For the orbital angular impulse, we shall derive the corresponding expression in Section \ref{sec7} as it is slightly more detailed.\\
	The KMOC formalism has been generalized to describe different types of scattering. For instance, in \cite{Cristofoli:2021vyo} the formalism has been extended to include incoming waves in the initial state. It has also been extended to include additional internal degrees of freedom like color charges in \cite{delaCruz:2020bbn}. Finally, it has been generalized to describe scattering in curved backgrounds \cite{Adamo:2021rfq, Adamo:2022rmp}. 
	\section{The Newman-Janis algorithm for three-point amplitudes} \label{NJ_algorithm}
	
	The Newman-Janis (NJ) algorithm has been known for a long time as a classical solution-generating technique, primarily used in the context of General relativity. In their original work \cite{nj}, Newman and Janis showed that one can ``derive'' the Kerr metric from the Schwarzschild solution (when written in the so-called Kerr-Schild coordinates) by doing a complex transformation of the radial coordinate, with the parameter by which it transforms interpreted as the spin of the Kerr black hole solution. Interestingly, they observed that there exists a similar mapping between solutions of the free Maxwell's equations as well. In electrodynamics, the NJ algorithm generates the so-called $\sqrt{\text{Kerr}}$ field from the Coulombic field of charged point particle sitting at the origin \cite{Lynden-Bell:2002dvr}.\\
	For static electromagnetic fields in vacuum, one can define the magnetostatic potential exactly as done for electrostatic solutions, since $\vec{\nabla} \times \vec{B} = 0\ \Rightarrow B_{i} = \partial_{i} \chi$. For a static point charge at the origin, then we have
	\begin{equation}
		\Phi (\vec{x})= \phi + i\chi = \frac{Q}{r},
	\end{equation}
	where $\phi$ and $\chi$ are the electrostatic and magnetostatic potential, respectively. Now, just as was done for the Kerr solution in GR, we do a complex transformation on the radial coordinate. We get
	\begin{equation}
		\Phi(\vec{x}) = \frac{Q}{r} \rightarrow \frac{Q}{\sqrt{(\vec{x}-i\vec{a})^{2}}} = \phi +i\chi.
	\end{equation}
	Here $\vec{a}$ is to be interpreted as the ring radius of the field, the radius at which there is a ring singularity. From the above expression, we can compute the $\sqrt{\text{Kerr}}$ electromagnetic field \cite{Lynden-Bell:2002dvr},
	\begin{equation}
		\vec{F} = -\vec{\nabla} \Phi = \vec{E} + i\vec{B}
	\end{equation}
	In the recent past, there have been investigations in understanding the Newman-Janis algorithm in effective field theory (EFT). As shown in \cite{Guevara:2020xjx}, the $\sqrt{\text{Kerr}}$ field in EFT can be thought of as being generated by a conserved current. The conserved current then defines a classical $\sqrt{\text{Kerr}}$ point particle with an infinite number of multipole moments described solely in terms of its mass ($m$), charge ($q$) and spin ($a$) \cite{Chung:2019yfs}. From the conserved current, we can compute the gauge field created by this configuration \cite{A:2022wsk},
	\begin{equation}
		A^{\mu}(x) = \int d^{4}x'\ G_{r}(x,x')\ J^{\mu}(x') = \sum _{n\geq 0} D^{\mu}_n (m,q,a) \frac{1}{r^n}
	\end{equation}
	where $D_{n}^{\mu}(m,q,a)$ are the multipole moments. This is the electromagnetic analog of the Kerr black hole, studied in \cite{Vines:2017hyw}. Remarkably, this recent understanding of $\sqrt{\text{Kerr}}$ field as a particle can also be understood, within the EFT framework \cite{Guevara:2020xjx}, as the classical limit of the three-point amplitude of a massive spin - $S$ particle interacting with a photon. We shall review this now.
	
	Consider the three-point amplitude for a generic massive spin - $S$ particle of mass $m_2$ ``minimally''\footnote{Here the term ``minimally'' means that the three-point amplitude is well behaved in the high energy limit \cite{Arkani-Hamed:2017jhn}.} coupled to a photon. In the massive spinor helicity formalism \cite{Arkani-Hamed:2017jhn}, the amplitude is given by
	\begin{align}\label{35}
		\mathcal{A}_3[\textbf{2}^S,\textbf{2}'^{S},q^{+}]&= \ i\sqrt{2}Q_{2} \ x \frac{\braket{\mathbf{2}\mathbf{2'}}^{2S}}{m_{2}^{2S-1}}\,,\cr 
		\mathcal{A}_3[\textbf{2}^S,\textbf{2}'^{S},q^{-}]&= i\sqrt{2}Q_{2}x^{-1}\frac{[\mathbf{2}\mathbf{2'}]^{2S}}{m_{2}^{2S-1}} \,.
	\end{align}
	Here the massive spinor helicity variables $|\textbf{2}\rangle$ and $|\textbf{2}'\rangle$ are defined w.r.t incoming momentum $p_2$ and outgoing momentum $p'_2$, respectively. The $x-$factor, which is a hallmark of a minimally coupled amplitude is defined via the photon polarization:
	$x=\frac{1}{m_2} (\varepsilon ^+ (q) \cdot p_2)$. 
	We now take the classical limit as described in the previous section. For the above amplitude, we replace $q=\hbar \bar{q}$ and $Q_{2}\rightarrow Q_{2}/\sqrt{\hbar}$. From the three particle kinematics, $2p_2 \cdot \bar{q} =- \hbar\bar{q}^2 $, keeping the leading term in $\hbar$, we obtain
	\begin{align}
		\frac{1}{m_2} \langle \mathbf{2} \mathbf{2}^\prime \rangle = \mathbb{I} + \frac{1}{2Sm_2}\bar{q} \cdot s_2 +\mathcal{O}(\hbar) \,,
	\end{align}
	where we have suppressed all the SU(2) indices. Here $s_2^\mu$ is the Pauli-Lubanski pseudovector associated with the spin-$S$ particle
	\begin{equation}\label{spin_vec_spinorreps}
		s_{2}^{\mu} = \frac{S\hbar}{m_{2}}\langle \mathbf{2} | \sigma^{\mu} | \mathbf{2}]\,.
	\end{equation}
	It was shown in \cite{Arkani-Hamed:2019ymq} that if we take $S\rightarrow \infty,\hbar\rightarrow 0$ such that $S\hbar=\text{constant}$, then the above three-point amplitude exponentiates 
	\begin{align}\label{47}
		\mathcal{A}_{3,\rootkerr}^{\pm} &= iQ_{2}\sqrt{2}m_{2} \ x^{\pm 1}e^{\pm \bar{q}\cdot a_{2}} \,,\cr
		&= \mathcal{A}^{\pm}_{3,\text{scalar}}\ e^{\pm \bar{q}\cdot a_{2}},
	\end{align}
	where $a_{2}^{\mu} = \frac{s_{2}^{\mu}}{m_2}$ is the rescaled spin of the $\sqrt{\text{Kerr}}$ particle. We note that the classical limit of the massive spin - $S$ particle has thus ``spun'' the three-point amplitude of a minimally coupled scalar (in the classical limit). This exponentiation is the realization of the Newman - Janis algorithm for three-point amplitudes, for scalars minimally coupled to the photon.		
	\section{Spin dressing of the photon propagator}\label{sec:lin_imp}
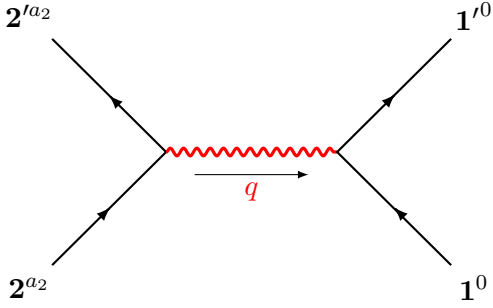
\begin{figure}[h!]
		\centering
		\begin{tikzpicture}[scale=1.5,photon/.style={decorate,decoration={snake, amplitude=0.5mm,segment length=1.75mm}}]
			\draw[thick] (-1,-1) -- (0,0) -- (-1,1);
			\draw[photon,very thick, red] (0,0) -- (1.5,0);
			\draw[thick] (2.5,-1) -- (1.5,0) -- (2.5,1);
			\draw[-latex, thick] (-0.5,-0.5) -- (-0.49,-0.49);
			\draw[-latex, thick] (-0.49,0.49) -- (-0.5,0.5);
			\draw[-latex, thick] (2,0.5) -- (2.01,0.51);
			\draw[-latex, thick] (2.01,-0.5) -- (2,-0.49);
			\draw[-latex] (0.25,-0.2) -- (1.25,-0.2) ;
			\node[red] at (0.75,-0.35){$q$};
			\node[black] at (-1.2,-1.2) {$\mathbf{2}^{a_2}$};
			\node[black] at (-1.2,1.2) {$\mathbf{2}^{\prime a_2}$};
			\node[black] at (2.7,-1.2) {$\mathbf{1}^0$};
			\node[black] at (2.7,1.2) {$\mathbf{1'}^{ 0}$};
		\end{tikzpicture}
		\caption{The four-point scalar-$\sqrt{\text{Kerr}}$ amplitude with photon exchange. Here `$a_{2}$' denotes the rescaled spin of the $\sqrt{\text{Kerr}}$ particle.}
		\label{linearimpulse}
	\end{figure}

	In this section, we interpret the exponentiation of the three-point amplitude, in eq.\eqref{47}, as a ``spin dressing'' of the photon propagator. 
	This is motivated by the simple observation that the three-point amplitude in eq.\eqref{47} can be written as
	\begin{equation}
		\begin{split}
			\mathcal{A}_{3,\rootkerr}^{\pm} &= iQ_{2}m_{2}\ (\varepsilon^{\pm}(\bar{q})\cdot p_{2})\ e^{\pm \bar{q}\cdot a_{2}},\\
			&= iQ_{2}m_{2} (\varepsilon'^{\pm}(\bar{q},a_{2})\cdot p_{2})
		\end{split}
	\end{equation}
	where $\varepsilon'^{\mu\pm}(\bar{q},a_{2}) = \varepsilon^{\mu\pm}(\bar{q}) e^{\pm a_{2}\cdot \bar{q}}$. This was first observed in \cite{Alessio:2023kgf}. Building on this observation,   
	we move on to the construction of
	the four-point scattering amplitude involving a scalar particle of charge $Q_{1}$ and mass $m_{1}$ and a $\rootkerr$ particle, mediated by photon. The incoming momenta for the  particles are $(p_1,p_2)$ and the outgoing momenta are $(p_1+q,p_2-q)$.
	Diagramatically, the four-point amplitude is represented in Figure \ref{linearimpulse}.
	The amplitude is then 
	\begin{align}\label{51}
		\mathcal{A}_4 \left[p_1,p_2 \rightarrow p'_1,p'_2 \right]  &= \mathcal{A}_{3,\rootkerr }^\mu \left[p'_2,p_2,\hbar\bar{q}  \right]\frac{\mathcal{P}_{\mu\nu}}{\hbar^{2}\bar{q}^{2}}\mathcal{A}_{3,\text{scalar}}^\nu \left[p'_1,p_1,-\hbar\bar{q}  \right] \,,
	\end{align}
	where $p'_1=p_1+\hbar \bq$, $p'_2=p_2-\hbar \bq$ and $\mathcal{P}^{\mu\nu} :=\sum _{h=\pm} \varepsilon _h^\mu\ \varepsilon^{\nu} _{-h}$. 
	Using the three-point amplitudes in eq.\eqref{47}, we note that eq.\eqref{51} can be written as
	\begin{align}
		\mathcal{A}_4\left[p_1,p_2 \rightarrow p'_1,p'_2 \right] &= \mathcal{A}_{3,\text{scalar}}^\mu \left[p'_2,p_2,\hbar\bar{q}  \right] \frac{\widetilde{\mathcal{P}}_{\mu\nu}(\bar{q})}{\hbar^{2}\bar{q}^{2}} \mathcal{A}_{3,\text{scalar}}^\nu \left[p'_1,p_1,\hbar\bar{q}  \right] \,, \label{alternatefourpoint}
	\end{align}
	where $\mathcal{A}_{3,\text{scalar}}^\mu$ is the three-point minimally coupled scalar amplitude. We now deform the internal photon projector $\mathcal{P}^{\mu \nu}$ as follows
	\begin{align}
		\mathcal{P}^{\mu \nu} \rightarrow   \widetilde{\mathcal{P}}^{\mu\nu}(\bar{q}) &:= e^{\bar{q} \cdot a_2} \varepsilon_{+}^\mu(\bar{q}) \varepsilon_{-}^\nu (\bar{q}) + e^{-\bar{q} \cdot a_2} \varepsilon_{+}^\nu(\bar{q}) \varepsilon_{-}^\mu (\bar{q}) \cr
		&= \cosh(a_2 \cdot \bar{q})\eta^{\mu\nu} + \sinh(a_2 \cdot \bar{q})\Pi^{\mu\nu} (\bar{q}) \,. \label{deformedprojector}
	\end{align}
	Here we have used $\varepsilon^{(\mu}_{+}\varepsilon_{-}^{\nu)} =\eta ^{\mu \nu}$ and define the anti-symmetric part of the projector as $\Pi^{\mu\nu} (\bar{q}) :=\varepsilon^{[\mu}_{+}\varepsilon_{-}^{\nu]}(\bq)$.  Since the anti-symmetric part\footnote{Usually the anti-symmetric part does not appear in the projector. In this case, we get it due to the helicity dependence of the exponentiation of the massive spin - $S$ amplitude in eq.\eqref{47}.}  of the projector is ambiguous up to a residual gauge, we shall choose an expression for $\Pi^{\mu \nu} (\bar{q})$ that can be used in the computation of all the physical observables. We choose \footnote{A similar construction of spin dressed photon propagator was obtained in \cite{Alessio:2023kgf}.}
	\begin{align}
		\Pi^{\mu\nu}(\bar{q}) =\frac{ i}{(a_2 \cdot \bar{q})}\epsilon^{\mu\nu \rho \sigma}a_{2\rho} \bar{q}_{\sigma} \,,
	\end{align}
	and substitute in equation eq.\eqref{alternatefourpoint} to obtain the amplitude 
\begin{align}
	\mathcal{A}_{4,\text{scalar} -\rootkerr } &=\frac{4  Q_{1}Q_{2}}{\hbar^2 \bar{q}^2} \left[ (p_1\cdot p_2) \cosh (\bar{q}\cdot a_2)  + i\frac{\sinh (\bar{q}\cdot a_{2})}{(\bar{q}\cdot a_{2})} \epsilon (p_1,p_2,a_2,\bar{q})\right]\,, \label{linearimpulseamplitude}
\end{align}
with $\epsilon (p_1,p_2,a_2,\bar{q}):=\epsilon _{\mu \nu \rho \sigma}p_1^\mu p_2^\nu a_2^\rho \bar{q}^\sigma $. The amplitude depends on the external momenta of the scattering particles as well as the (classical) spin vector $a^{\mu}$. It is related to the spin tensor $S^{\mu\nu}$ via the dual relation, eq.\eqref{11}. It is rather natural to interpret $S^{\mu\nu}$ as the independent spin tensor which can be thought of as an ``intrinsic'' spin angular momentum of a classical particle. In this case 
\begin{align}\label{47s}
	a_2^{\mu}\, =\, a_2^{\mu}(S_2, p_2) \,.
\end{align}
We will denote the projection of $S_{2}^{\mu\nu}$ orthogonal to the time-like vector $p_{2}^\mu$ as $S_{2}^{\perp \mu\nu}$. Thus we will interpret the spin pseudovector as a function of $S_{2}^{\perp\mu\nu}$ and $p_2^\mu$. We will not explicitly indicate the dependence of $a_{2}^\mu$ on $S_{2}^{\mu\nu}$ except in section \ref{sec7}, when we derive the angular impulse.

Using the above amplitude, the linear impulse for the scalar particle is
\begin{equation}\label{lin_imp_scalar_kmo}
	\begin{split}
		\Delta p_1^\mu &= iQ_{1}Q_{2} \int \hat{d}^4 \bar{q}\ \hat{\delta}(\bar{q}\cdot p_1)\ \hat{\delta}(\bar{q} \cdot p_2)\frac{e^{i \bar{q} \cdot b}}{\bar{q}^2}\ \bar{q}^\mu \\ 
		&\hspace{3cm}\Big[ \cosh(a_2 \cdot \bar{q})\ (p_1 \cdot p_2) + i \frac{\sinh (a_2 \cdot \bar{q})}{(a_2 \cdot \bar{q})} \epsilon(p_1,p_2, a_2,\bar{q}) \Big]\, 
	\end{split}
\end{equation}
Using the identities
\begin{align}
	\bar{q}_{\mu}\sinh{w} = i\epsilon_{\mu\nu\rho\sigma}u_{1}^{\nu}u_{2}^{\rho}\bar{q}^{\sigma} \,,\ \ \sinh w = \sqrt{\gamma^{2}-1}\,, \quad \gamma =(u_1\cdot u_2)\,, \label{gramidentity}
\end{align}
and rewriting the $\cosh (a_{2}\cdot \bar{q})$ and $\sinh (a_{2}\cdot \bar{q})$ terms as exponential functions, we obtain
\begin{equation}
	\Delta p_{1}^{\mu} = \frac{Q_{1}Q_{2}}{2\pi\gamma\beta} ~ \text{Re}~ \Big[\frac{\gamma(b+i\Pi a_2)^{\mu} - i\epsilon^{\mu}(b+i\Pi a_2,u_1,u_2)}{(b+i\Pi a_2)^2} \Big].
\end{equation}
This is the expression obtained in \cite{Arkani-Hamed:2019ymq}. 
Hence, we see that the NJ algorithm, within the EFT for a $\sqrt{\text{Kerr}}$ particle, can also be interpreted as a deformation on the photon data rather than on the impact parameter as shown in \cite{Arkani-Hamed:2019ymq}.

\section{Radiation Kernel to all order in spin}\label{rad_ker}

\begin{figure}[h!]
	\centering
	\begin{tikzpicture}[scale=1,photon/.style={decorate,decoration={snake, amplitude=0.5mm,segment length=1.75mm}}]
		\filldraw [black] (0,0) circle (3mm);
		\draw[thick] (-1,-1) -- (0,0);
		\draw[thick] (-1,1) -- (0,0);
		\draw[-latex, thick] (-0.7,-0.7) -- (-0.71,-0.71);
		\draw[-latex, thick] (-0.51,0.51) -- (-0.5,0.5);
		\draw[thick] (1,1) -- (0,0);
		\draw[thick] (1,-1) -- (0,0);
		\draw[-latex, thick] (0.51,0.51) -- (0.5,0.5);
		\draw[-latex, thick] (0.7,-0.7) -- (0.71,-0.71);
		\draw[->,photon, thick, black] (0.3,0) -- (1,0);
		\node[black] at (-1.2,-1.2) {$\mathbf{2}^{a_2}$};
		\node[black] at (-1.2,1.2) {$\mathbf{\bar{2}}^{a_2}$};
		\node[black] at (1.2,1.2) {$\mathbf{\bar{1}}^{0}$};
		\node[black] at (1.2,-1.2) {$\mathbf{1}^{0}$};
		\node[black] at (1.2,0) {$k$};
	\end{tikzpicture}
	\caption{The five-point amplitude appearing in the radiation
		kernel at leading order in coupling.}
	\label{5ptradkernel}
\end{figure}
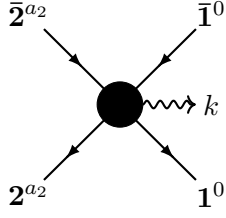

\begin{figure}
	\centering
	\begin{tikzpicture}[scale=1,photon/.style={decorate,decoration={snake, amplitude=0.5mm,segment length=1.75mm}}]
		\draw[thick] (-1,-1) -- (0,0) -- (-1,1);
		\draw[photon, thick, red] (0,0) -- (1.5,0);
		\draw[->,photon, thick] (2,0.5) -- (1.5,1);
		\draw[thick] (2.5,-1) -- (1.5,0) -- (2.5,1);
		\draw[-latex, thick] (-0.5,-0.5) -- (-0.51,-0.51);
		\draw[-latex, thick] (-0.5,0.5) -- (-0.49,0.49);
		\draw[-latex, thick] (2.2,0.7) -- (2.19,0.69);
		\draw[-latex, thick] (2,-0.5) -- (2.01,-0.51);
		\draw[-latex] (0.25,-0.2) -- (1.25,-0.2) ;
		\node[red] at (0.75,-0.35){$q_2$};
		\node[black] at (-1.2,-1.2) {$\mathbf{2}$};
		\node[black] at (-1.2,1.2) {$\mathbf{\bar{2}}$};
		\node[black] at (2.7,-1.2) {$\mathbf{1}$};
		\node[black] at (2.7,1.2) {$\mathbf{\bar{1}}$};
		\node[black] at (1.4,1.2) {$k$};
		\node[] at (1,-1.5) {I};
	\end{tikzpicture}
	\quad
	\begin{tikzpicture}[scale=1,photon/.style={decorate,decoration={snake, amplitude=0.5mm,segment length=1.75mm}}]
		\draw[thick] (-1,-1) -- (0,0) -- (-1,1);
		\draw[photon, thick, red] (0,0) -- (1.5,0);
		\draw[->,photon, thick] (2,-0.5) -- (2.5,0);
		\draw[thick] (2.5,-1) -- (1.5,0) -- (2.5,1);
		\draw[-latex, thick] (-0.5,-0.5) -- (-0.51,-0.51);
		\draw[-latex, thick] (-0.5,0.5) -- (-0.49,0.49);
		\draw[-latex, thick] (2.01,0.51) -- (2,0.5);
		\draw[-latex, thick] (2.3,-0.8) -- (2.31,-0.81);
		\draw[-latex] (0.25,-0.2) -- (1.25,-0.2) ;
		\node[red] at (0.75,-0.35){$q_2$};
		\node[black] at (-1.2,-1.2) {$\mathbf{2}^{a_2}$};
		\node[black] at (-1.2,1.2) {$\mathbf{\bar{2}}^{a_2}$};
		\node[black] at (2.7,-1.2) {$\mathbf{1}^{0}$};
		\node[black] at (2.7,1.2) {$\mathbf{\bar{1}}^{0}$};
		\node[black] at (2.7,0.1) {$k$};
		\node[] at (1,-1.5) {II};
	\end{tikzpicture}
	\quad
	\begin{tikzpicture}[scale=1,photon/.style={decorate,decoration={snake, amplitude=0.5mm,segment length=1.75mm}}]
		\draw[thick] (-1,-1) -- (0,0) -- (-1,1);
		\draw[photon, thick, red] (0,0) -- (1.5,0);
		\draw[->,photon, thick] (1.5,0) -- (1,1);
		\draw[thick] (2.5,-1) -- (1.5,0) -- (2.5,1);
		\draw[-latex, thick] (-0.5,-0.5) -- (-0.51,-0.51);
		\draw[-latex, thick] (-0.5,0.5) -- (-0.49,0.49);
		\draw[-latex, thick] (2.01,0.51) -- (2,0.5);
		\draw[-latex, thick] (2,-0.5) -- (2.01,-0.51);
		\draw[-latex] (0.25,-0.2) -- (1.25,-0.2) ;
		\node[red] at (0.75,-0.35){$q_2$};
		\node[black] at (-1.2,-1.2) {$\mathbf{2}^{a_2}$};
		\node[black] at (-1.2,1.2) {$\mathbf{\bar{2}}^{a_2}$};
		\node[black] at (2.7,-1.2) {$\mathbf{\bar{1}}^{0}$};
		\node[black] at (2.7,1.2) {$\mathbf{\bar{1}}^{0}$};
		\node[black] at (0.9,1.2) {$k$};
		\node[] at (1,-1.5) {III};
	\end{tikzpicture}
	\caption{Diagrams contributing to the tree level five-particle amplitude with a photon emitted from the scalar particle.} \label{fig:compton}
\end{figure}
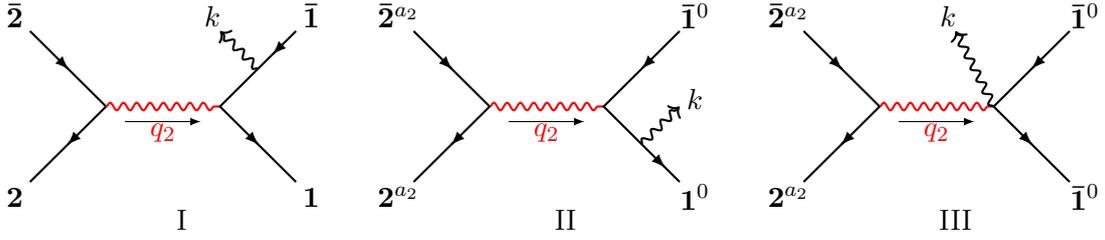

In this section, we use the spin-dressed photon propagator \eqref{deformedprojector} to compute the leading order radiative gauge field emitted by a scalar particle as it scatters in the background of a $\rootkerr$ particle. The basic ingredient for computing the radiative field via KMOC formalism is the inelastic five-point amplitude as shown in Figure \ref{5ptradkernel}. 
We will only compute the radiation emitted from the scalar particle using the NJ algorithm. It is straightforward to compute the five-point amplitude when the photon is emitted from the scalar particle since, in this case, the complexity due to the spin is completely contained within the three-point amplitude involving the $\rootkerr $ particle.  The other sub-amplitude needed to obtain the full amplitude is then the ordinary scalar-Compton amplitude as indicated in Figure \ref{fig:compton}.
Following the construction of the four-point massive amplitude in section \ref{sec:lin_imp}, we use the deformed internal photon projector \eqref{deformedprojector} 
\begin{align}
	\widetilde{\mathcal{P}}^{\mu \nu}(\bar{q}_2) =\cosh (a_2\cdot \bar{q}_2) \eta _{\mu \nu}  +i \frac{\sinh (a_2\cdot \bar{q}_2)}{a_2\cdot \bar{q}_2} \epsilon ^{\mu \nu \rho \sigma} a_{2\rho}\bar{q}_{2\sigma} \,.
\end{align}
to obtain the five-point amplitude as follows
\begin{align}
	\mathcal{A}^\delta _{5}&=\frac{1}{q_2^2} \mathcal{A}^\mu _{3,\rootkerr}  \mathcal{P}_{\mu \nu} \mathcal{A}_{4,\text{Scalar-Compton}}^{\nu \delta}\cr
	& =\frac{1}{q_2^2} \mathcal{A}^\mu _{3,\text{Scalar}}  \widetilde{\mathcal{P}}_{\mu \nu} \mathcal{A}_{4,\text{Scalar-Compton}}^{\nu \delta} \,.
\end{align} 
Here $ \mathcal{A}^\mu _{3,\text{Scalar}}$ and $\mathcal{A}_{4,\text{Scalar-Compton}}^{\nu \delta} $ are the three-point massive scalar-photon and the scalar-Compton amplitude in scalar QED.
To derive the amplitude, we'll be using the following momentum convention
\begin{align}
	(\bar{p}_1,\bar{p}_2)\rightarrow (p_1,p_2,k)\,, \quad \text{with} \quad \bar{p}_1=p_1+q_1\,, \quad \bar{p}_2=p_2+q_2\,, \quad k=(q_1+q_2)\,,
\end{align}
as indicated in Figure \ref{fig:compton}. There are three scattering diagrams. We start with diagram I in Figure \ref{fig:compton}. It is given by
\begin{align}
	\mathcal{A}^\delta _{5,I} &=4Q_{1}^{2}Q_{2} \frac{(2p_2 +q_2)^\mu  p_1^\nu( p_1-q_2)^\delta }{q_2^2[(p_1-q_2)^2-m_1^2]} \left[\cosh (a_2\cdot \bar{q}_2) \eta _{\mu \nu}  +\frac{ i \sinh (a_2\cdot \bar{q}_2)}{a_2\cdot \bar{q}_2} \epsilon ^{\mu \nu \rho \sigma} a_{2\rho}\bar{q}_{2\sigma}  \right]\cr
	&= \frac{4Q_{1}^{2}Q_{2} ( p_1-q_2)^\delta}{q_2^2 [(p_1-q_2)^2-m_1^2]} \left[\cosh (a_2\cdot \bar{q}_2) (2p_1\cdot p_2 +k\cdot p_1)  +2i \frac{ \sinh (a_2\cdot \bar{q}_2)}{a_2\cdot \bar{q}_2} \epsilon (p_2,p_1,a_2,\bar{q}_2)  \right]\,.\cr
\end{align}
Similarly, from diagram II, we obtain
\begin{align}
	\mathcal{A}^\delta _{5,II} =\frac{4Q_{1}^{2}Q_{2}}{q_2^2(k\cdot p_1)} p_1^\delta \Big[\cosh (a_2\cdot \bar{q}_2)& \left(p_1\cdot p_2 +k\cdot p_2+{\frac{k\cdot p_1}{2}+\frac{k\cdot q_2}{2}} \right) \cr
	& +\frac{ i \sinh (a_2\cdot \bar{q}_2)}{a_2\cdot \bar{q}_2} \epsilon (p_2,p_1+k,a_2,\bar{q}_2)  \Big]\,.
\end{align}
We use the usual Feynman rule for the scalar-photon four-point vertex in scalar QED theory to obtain the contribution from diagram III in Figure \ref{fig:compton}
\begin{align}
	\mathcal{A}_{5,III}^\delta &=-2Q_{1}^{2}Q_{2} \eta ^{\rho \delta} \frac{1}{q_2^2}  (2p_2+q_2)^\nu \left[\cosh (a_2\cdot \bar{q}_2) \eta _{\rho \nu}  +\frac{ i\sinh (a_2\cdot \bar{q}_2)}{a_2\cdot \bar{q}_2} \epsilon _{ \nu \rho \sigma \alpha} a^{\sigma}_2 \bar{q}^{\alpha}_2  \right] \,,\cr
	&=-4Q_{1}^{2}Q_{2}\frac{1}{q_2^2}\left[\cosh (a_2\cdot \bar{q}_2) \left( p_2+\frac{q_2}{2}\right)^\delta  -\frac{ i \sinh (a_2\cdot \bar{q}_2)}{a_2\cdot \bar{q}_2} \epsilon ^{ \delta} (p_2,a_2,\bar{q}_2)  \right]\,.
\end{align}

Next, we scale the massless momentum $q_2^\mu$ as $\hbar \bar{q}_2^\mu$ and collect the terms of $\mathcal{O}(\hbar^{-2})$ needed to compute the leading order radiation kernel. But unlike the four-point case,  individual diagrams in this tree-level amplitude contain superclassical terms. As expected, they cancel after summing up all the diagrams. To see this first of all we rewrite the massive propagator in diagram I as 
\begin{align}
	[(p_1-q_2)^2-m_1^2]^{-1} = -\frac{1}{2\hbar \bar{k}\cdot p_1} \left( 1-\hbar \frac{\bar{k}\cdot \bar{q}_2}{\bar{k}\cdot p_1}\right)^{-1}\,,
\end{align}
where we set $k^2=0$ and $p_2\cdot q_2 =-\frac{q_2^2}{2}$.  The second condition is due to one of the two on-shell delta functions present in the radiation kernel.
Expanding the contribution from diagram I upto $\mathcal{O}(\hbar ^{-2})$, we find
\begin{align}
	\mathcal{A}^\delta _{5,I,\mathcal{O}(\hbar^{-2})} = \frac{4Q_{1}^{2}Q_{2}  }{\hbar^2 \bar{q}_2^2(\bar{k}\cdot p_1)} \left[\cosh (a_2\cdot \bar{q}_2) (p_1\cdot p_2)  +i\frac{  \sinh (a_2\cdot \bar{q}_2)}{a_2\cdot \bar{q}_2} \epsilon (p_2,p_1,a_2,\bar{q}_2)  \right]\cr \times \left(\bar{q}_2^\delta -\frac{p_1^\delta (\bar{k}\cdot \bar{q}_2)}{\bar{k}\cdot p_1}  \right)-\frac{2Q_{1}^{2}Q_{2}  }{\hbar^2 \bar{q}_2^2}\cosh (a_2\cdot \bar{q}_2)p_1^\delta \,,
\end{align}
and 
\begin{align}
	\mathcal{A}^\delta _{5,I,\mathcal{O}(\hbar^{-3})} = \frac{4Q_{1}^{2}Q_{2}  p_1^\delta }{\hbar^3 \bar{q}_2^2(\bar{k}\cdot p_1)} \left[ (-p_1\cdot p_2)\cosh (a_2\cdot \bar{q}_2)-i\frac{  \sinh (a_2\cdot \bar{q}_2)}{a_2\cdot \bar{q}_2} \epsilon (p_2,p_1,a_2,\bar{q}_2) \right]
\end{align}
Similarly, from diagram II, we get both $\mathcal{O}(\hbar ^{-2})$ and $\mathcal{O}(\hbar ^{-3})$ terms but the latter cancels with the contribution from I.
\begin{align}
	\mathcal{A}^\delta _{5,II,\mathcal{O}(\hbar^{-2})} &=\frac{4Q_{1}^{2}Q_{2} p_1^\delta }{\hbar^2 \bar{q}_2^2(\bar{k}\cdot p_1)}\left[\cosh (a_2\cdot \bar{q}_2) \left(\bar{k}\cdot p_2 +\frac{\bar{k}\cdot p_1}{2}\right)  +i \frac{ \sinh (a_2\cdot \bar{q}_2)}{a_2\cdot \bar{q}_2} \epsilon (p_2,\bar{k},a_2,\bar{q}_2)  \right]\,,\cr
	\mathcal{A}^\delta _{5,II,\mathcal{O}(\hbar^{-3})} &= \frac{4Q_{1}^{2}Q_{2}  p_1^\delta }{\hbar^3 \bar{q}_2^2(\bar{k}\cdot p_1)} \left[ (p_1\cdot p_2)\cosh (a_2\cdot \bar{q}_2)+i\frac{  \sinh (a_2\cdot \bar{q}_2)}{a_2\cdot \bar{q}_2} \epsilon (p_2,p_1,a_2,\bar{q}_2) \right]
\end{align}
The $\mathcal{O}(\hbar^{-2})$ terms from diagram III can be found trivially. We collect all the terms of $\mathcal{O}(\hbar^{-2})$ below
\begin{align}
	\mathcal{A}^\delta _{5,\mathcal{O}(\hbar^{-2})} 
	=\frac{4Q_{1}^{2}Q_{2}}{\hbar ^2 \bar{q}_2^2}\frac{m_1m_2}{\bar{k}\cdot p_1} \Big[ \cosh (a_2\cdot \bar{q}_2)\big\lbrace \gamma  \bar{q}_2^\delta -u_2^\delta (\bar{k}\cdot u_1) 
	-\frac{p_1^\delta}{\bar{k}\cdot p_1}\left( \gamma (\bar{k}\cdot \bar{q}_2)-(\bar{k}\cdot u_2)(\bar{k}\cdot u_1)\right) \big\rbrace \cr
	+i\sinh (a_2\cdot \bar{q}_2) \big\lbrace \epsilon ^\delta (u_2,u_1,\bar{q}_2)-\frac{p_1^\delta}{\bar{k}\cdot p_1} \epsilon (\bar{k},u_2,u_1,\bar{q}_2)\big\rbrace
	\Big]\,.\label{radker_scala_kmo_exacta}
\end{align}

Using the formula of eq. \eqref{radiationkernel}, we obtain the radiation kernel as
\begin{align}\label{radker_scala_kmo_exacta1}
	\mathcal{R}^{\mu}_1 (\bar{k},a_2) &=Q_{1}^{2}Q_{2} \int \hat{d}^4 \bq \hat{\delta}[u_1\cdot (\bq-\bar{k})] \hat{\delta}(u_2\cdot \bq) \frac{e^{-i\bq\cdot b}}{\bq^2}\frac{1}{\bar{k}\cdot p_1}\cr 
	& \times\Big[  \cosh(a_2\cdot \bq) \lbrace \gamma \bq^\mu -u_2^\mu (u_1\cdot \bar{k})\rbrace+i\sinh(a_2\cdot \bq)\epsilon ^{\mu}(u_2,u_1,\bq)  \cr
	& -\frac{p_1^\mu}{\bar{k}\cdot p_1}\big\lbrace \cosh (a_2\cdot \bq)\left( \gamma (\bar{k}\cdot \bq)- (\bar{k}\cdot u_2) (u_1\cdot \bar{k})\right)  + i\sinh (a_2\cdot \bq) \epsilon (\bar{k},u_2,u_1,\bq)  \big \rbrace \Big]\,.\cr
\end{align}
This expression agrees with the result in \eqref{radker_scalar_exacta}, which is obtained using classical equations of motion.\\
The radiation kernel for the scalar can also be implemented as a complex shift in the impact parameter space, just as it was done for linear impulse in \cite{Arkani-Hamed:2019ymq}.
Consider the radiation kernel for scalar-scalar scattering in electrodynamics 
\begin{align}
	\mathcal{R}^{\mu}_1 (\bar{k}) =Q_{1}^{2}Q_{2} \int  \hat{d}^4 \bar{q} & \hat{\delta}(u_2\cdot \bar{q}) \hat{\delta}[u_1\cdot (\bar{q}-\bar{k})]  \frac{e^{-i\bar{q}\cdot b}}{\bar{q}^2}\frac{1}{\bar{k}\cdot u_1}\cr 
	& \times\left[\bar{k}_\alpha(u_{1}\wedge \partial _{u_1})^{\mu\alpha} (u_1\cdot u_2) +(u_1\cdot u_2) \left \lbrace \bar{q}^\mu -\frac{(\bar{k}\cdot \bar{q})u_1^\mu}{\bar{k}\cdot u_1} \right\rbrace \right]\,.
\end{align}
Inside this expression for the radiation kernel, we complexify $\mathfrak{b}=b+ia_2$ and re-write the expression as follows.
\begin{align}
	\mathcal{R}^{\mu}_1 (\bar{k},a_2)& =Q_{1}^{2}Q_{2} \int  \hat{d}^4 \bq \hat{\delta}(u_2\cdot \bq) \hat{\delta}[u_1\cdot (\bq-\bar{k})]  \frac{1}{\bq^2(\bar{k}\cdot u_1)}\cr 
	& \times\Big[k_\alpha(u_1 \wedge \partial _{u_1})^{\mu \alpha} (e^{-i \bar{q} \cdot \mathfrak{b}^\star}e^w + e^{-i \bar{q} \cdot \mathfrak{b}}e^{-w})  +(e^{-i \bar{q} \cdot \mathfrak{b}^\star}e^w + e^{-i \bar{q} \cdot \mathfrak{b}}e^{-w})\left \lbrace \bq^\mu -\frac{(\bar{k}\cdot \bq)u_1^\mu}{\bar{k}\cdot u_1} \right\rbrace \Big]\,,\cr
\end{align}
with $u_1\cdot u_2 =\cosh w =\gamma$.
We can now factor out the overall $e^{-i \bar{q}\cdot b}$ in the second line
\begin{align}
	(e^{-i \bar{q} \cdot \mathfrak{b}^\star}e^w + e^{-i \bar{q} \cdot \mathfrak{b}}e^{-w})=e^{-i\bar{q}\cdot b} \left[ \gamma \cosh (\bq \cdot a_2)-\sqrt{\gamma^{2}-1}\sinh (\bq\cdot a_2)\right]\,,
\end{align}
where we use $\gamma = \cosh w$ and $\sqrt{\gamma^{2}-1} =\sinh w$.
We can now evaluate the derivatives with respect to initial velocity $u_{1}$,
\begin{align}
	&(u_{1}\wedge \partial _{u_1})^{\mu\nu} \left[ \gamma \cosh (\bq\cdot a_2)-\sqrt{\gamma^{2}-1}\sinh (\bq\cdot a_2)\right]\cr
	& =(u_1\wedge u_2)^{\mu \nu} \left[\cosh (a_2\cdot \bq)- \frac{1}{\beta}\sinh (a_2\cdot \bq) \right] \,,
\end{align} 
with $\sinh w =\beta \gamma$.  Using the following identity, consistent with the two on-shell delta function constraints: $u_2\cdot \bar{q} =0$ and $u_1\cdot \bar{q} =\bar{k}\cdot u_1$, we find that
\begin{align}
	\bq^\mu \sinh w =\frac{\bar{k}\cdot u_1}{\sinh w}(\gamma u_2^\mu -u_1^\mu) +i\epsilon ^\mu (\bq,u_1,u_2)\,,
\end{align}
and we recover the radiation kernel in scalar-$\rootkerr$ scattering when the photon is emitted from the scalar particle in \eqref{radker_scala_kmo_exacta1}. 

\begin{figure}
	\centering
	\begin{tikzpicture}[scale=1,photon/.style={decorate,decoration={snake, amplitude=0.5mm,segment length=1.75mm}}]
		\draw[thick] (-1,-1) -- (0,0) -- (-1,1);
		\draw[photon, thick, red] (0,0) -- (1.5,0);
		\draw[->,photon, thick] (-0.5,0.5) -- (0,1);
		\draw[thick] (2.5,-1) -- (1.5,0) -- (2.5,1);
		\draw[-latex, thick] (-0.5,-0.5) -- (-0.51,-0.51);
		\draw[-latex, thick] (-0.6,0.6) -- (-0.59,0.59);
		\draw[-latex, thick] (2.01,0.51) -- (2,0.5);
		\draw[-latex, thick] (2,-0.5) -- (2.01,-0.51);
		\draw[-latex] (1.25,-0.2) -- (0.25,-0.2) ;
		\node[red] at (0.75,-0.35){$q_1$};
		\node[black] at (-1.2,-1.2) {$\mathbf{2}^{a_2}$};
		\node[black] at (-1.2,1.2) {$\mathbf{\bar{2}}^{a_2}$};
		\node[black] at (2.7,-1.2) {$\mathbf{1}^{0}$};
		\node[black] at (2.7,1.2) {$\mathbf{\bar{1}}^{0}$};
		\node[black] at (0.2,1.2) {$k$};
		\node[] at (1,-1.5) {I};
	\end{tikzpicture}
	\quad
	\begin{tikzpicture}[scale=1,photon/.style={decorate,decoration={snake, amplitude=0.5mm,segment length=1.75mm}}]
		\draw[thick] (-1,-1) -- (0,0) -- (-1,1);
		\draw[photon, thick, red] (0,0) -- (1.5,0);
		\draw[->,photon, thick] (-0.5,-0.5) -- (-1,0);
		\draw[thick] (2.5,-1) -- (1.5,0) -- (2.5,1);
		\draw[-latex, thick] (-0.8,-0.8) -- (-0.81,-0.81);
		\draw[-latex, thick] (-0.5,0.5) -- (-0.49,0.49);
		\draw[-latex, thick] (2.01,0.51) -- (2,0.5);
		\draw[-latex, thick] (2,-0.5) -- (2.01,-0.51);
		\draw[-latex] (1.25,-0.2) --(0.25,-0.2)  ;
		\node[red] at (0.75,-0.35){$q_1$};
		\node[black] at (-1.2,-1.2) {$\mathbf{2}^{a_2}$};
		\node[black] at (-1.2,1.2) {$\mathbf{\bar{2}}^{a_2}$};
		\node[black] at (2.7,-1.2) {$\mathbf{1}^{0}$};
		\node[black] at (2.7,1.2) {$\mathbf{\bar{1}}^{0}$};
		\node[black] at (-1.1,0.2) {$k$};
		\node[] at (1,-1.5) {II};
	\end{tikzpicture}
	\quad
	\begin{tikzpicture}[scale=1,photon/.style={decorate,decoration={snake, amplitude=0.5mm,segment length=1.75mm}}]
		\draw[thick] (-1,-1) -- (0,0) -- (-1,1);
		\draw[photon, thick, red] (0,0) -- (1.5,0);
		\draw[->,photon, thick] (0,0) -- (1,1);
		\draw[thick] (2.5,-1) -- (1.5,0) -- (2.5,1);
		\draw[-latex, thick] (-0.5,-0.5) -- (-0.51,-0.51);
		\draw[-latex, thick] (-0.5,0.5) -- (-0.49,0.49);
		\draw[-latex, thick] (2.01,0.51) -- (2,0.5);
		\draw[-latex, thick] (2,-0.5) -- (2.01,-0.51);
		\draw[-latex] (1.25,-0.2) -- (0.25,-0.2) ;
		\node[red] at (0.75,-0.35){$q_1$};
		\node[black] at (-1.2,-1.2) {$\mathbf{2}^{a_2}$};
		\node[black] at (-1.2,1.2) {$\mathbf{\bar{2}}^{a_2}$};
		\node[black] at (2.7,-1.2) {$\mathbf{\bar{1}}^{0}$};
		\node[black] at (2.7,1.2) {$\mathbf{\bar{1}}^{0}$};
		\node[black] at (1.1,1.2) {$k$};
		\node[] at (1,-1.5) {III};
	\end{tikzpicture}
	\caption{Diagrams contributing to the tree level five-point amplitude with a photon emitted from the $\rootkerr$ particle.} \label{fig:comptonkerr}
\end{figure}
We note that to compute the total radiative flux (which includes the radiation emitted by the $\rootkerr$ particle), we require the expression for the five-point amplitude where the incoming $\rootkerr$ and scalar states are scattered into $\rootkerr$, scalar and a photon. A diagrammatic representation of this amplitude is in Fig. \ref{fig:comptonkerr}.

One avenue to compute the inelastic amplitude ${\cal A}_{5}(\rootkerr + \textrm{scalar}\, \rightarrow\, \rootkerr + \textrm{scalar} + \gamma)$ is via an EFT computation of the Compton sub-amplitude which has been pursued extensively in the literature recently \cite{Chung:2018kqs,Guevara:2018wpp,Guevara:2017csg,Alessio:2023kgf,Aoude:2020onz,Haddad:2020tvs,Chiodaroli:2021eug,Aoude:2022trd,Bern:2022kto,Aoude:2022thd,Chen:2022clh,Cangemi:2022abk,Saketh:2022wap,Bjerrum-Bohr:2023jau,Bjerrum-Bohr:2023iey,Haddad:2023ylx,Brandhuber:2023hhy,Aoude:2023vdk,Bern:2023ity,Brandhuber:2023hhl, Scheopner:2023rzp,Bautista:2021wfy,Bautista:2022wjf,Bautista:2023szu,Bautista:2023sdf}. Hence a possible strategy to compute the leading order radiation kernel in the present case would be to use the Compton sub-amplitude to evaluate all the diagrams in Fig. \ref{fig:comptonkerr}.  

However, a more natural route is to start with a gauge invariant bare Lagrangian of QED with $\rootkerr$ charged matter and compute the five-point amplitude directly using the resulting Feynman rules. Given that the three-point coupling of $\rootkerr$ with a photon is known, one can in principle use gauge invariance to fix all the higher point couplings. In 
\cite{Alessio:2023kgf}, it was shown that just as in the case of scalars and fermions, a gauge-invariant Lagrangian which describes the minimal coupling of a  $\rootkerr$  particle with the electromagnetic field is simply the scalar QED lagrangian in which the gauge covariant derivative is replaced by a twisted covariant derivative, $D_\mu^{(a)} = \partial_\mu - i e \exp{ \{\epsilon^\nu {}_\mu (a,\partial)\}} A_\nu$. It would be rather natural to simply use this Lagrangian and compute the radiative field emitted during scalar$-\rootkerr$ scattering. However as the author emphasizes in \cite{Alessio:2023kgf}, such a Lagrangian is not consistent with all the Compton amplitudes, and as a result, it is unclear how it would lead to the correct answer for classical radiation. We stress that the computation of the complete radiation kernel emitted during scalar$-\rootkerr$ scattering using the KMOC formalism has the potential to unravel the full power of the NJ algorithm.  This will be pursued elsewhere \cite{Akhtar:2024rzp}.


\section{Leading order angular impulse} \label{sec7}

In this section, we use the NJ algorithm to compute the leading order angular impulse $\Delta J^{\mu\nu}$ for the scalar and $\rootkerr$ particles. Angular impulse is an intriguing observable in $D = 4$ dimensions. Already for the scattering of scalar particles, it was shown in \cite{Gralla:2021eoi} that the net angular impulse of the particles (in a $ 2 \rightarrow\, 2$ scattering) does not add up to zero even at leading order in the coupling. The missing contribution is due to angular momentum stored in the late-time Coloumbic modes. In \cite{Gralla:2021eoi} this contribution was called electromagnetic scoot.  We will denote the scoot as $\delta^{\mu\nu}_{\textrm{scalar-scoot}}$, where the subscript indicates that the scoot has been computed for the case of scalar-scalar scattering. 

The NJ algorithm offers a powerful tool to compute angular impulse for the scalar-$\rootkerr$ system. We will denote the angular impulse for the scalar particle as $\Delta L^{\mu\nu}_{1}$. It can be computed to all orders in spin using the NJ algorithm and the final result is given in eq.\eqref{orb_imp_1_kmo}. The computation of total angular impulse (i.e. change in orbital angular momentum, $\Delta L_2^{\mu\nu}$  plus the change in spin angular momentum, $\Delta S_2^{\perp\mu\nu}$) for the $\rootkerr$ particle can also be done using the NJ algorithm. We will denote it as $\Delta J_{2}^{\mu\nu}$. As noted below eq.\eqref{47s}, for the computation of the orbital angular impulse for the $\rootkerr$ particle, we take $S_{2}^{\perp\mu\nu}$ as the independent variable and take $a_{2}^{\mu}(S^{\perp}_{2},p_{2})$\footnote{A moment of reflection reveals that for orbital angular impulse of $\rootkerr$ particle, the choice of $a_2^\mu$ versus $S_2^{\mu\nu}$ as independent variable in ${\cal A}_{4}$ will produce inequivalent results as $\Ll_{2}^{\mu\nu} =i \hbar (p_{2}\, \wedge\ \partial_{p_{2}})^{\mu\nu}$.}. A result (in the integral form) for $\Delta L_2^{\mu\nu}$ and $\Delta S_{2}^{\perp\mu\nu}$ appears in eq.\eqref{ang_imp_KMOC2} and \eqref{spinangularallorder}, respectively.  In principle, this completes the computation of angular impulse for the scalar-$\rootkerr$ system to leading order in coupling. 

To test our results, we compute the net angular impulse of the scattering particles and subject it to the conservation law.

Based on \cite{Gralla:2021eoi} we deduce that to leading order in the coupling, 
\begin{align}\label{totalangcons}
	\Delta L^{\mu\nu}_{1} + \Delta J^{\mu\nu}_{2} + \delta^{\mu\nu}_{\textrm{scoot}}\, =\, 0
\end{align}
On general grounds, we expect that the entire contribution to the electromagnetic scoot is independent of the spin of the particles as it simply arises due to the late-time Coloumbic effects which do not depend on the spin, 
\begin{align}
	\delta^{\mu\nu}_{\textrm{scoot}}\, =\, \delta^{\mu\nu}_{\textrm{scalar-scoot}} \,.
\end{align}
We thus expect that 
\begin{align}\label{totalangconsss}
	\Delta L^{\mu\nu}_{1} + \Delta J^{\mu\nu}_{2} + \delta^{\mu\nu}_{\textrm{scalar-scoot}}\, =\, 0 \,.
\end{align}
We verify the conservation to next to leading (i.e. linear) order in $S_{2}^{\mu\nu}$ in a perturbative expansion which is valid when $\vert a_2 \vert \ll \vert b \vert$. As we will argue, verifying conservation for finite spin $\vert a_2 \vert \sim\, \vert b\vert$ is rather subtle and will be pursued elsewhere \cite{Akhtar:2024rzp}. \\
We start with the computation of the orbital angular impulse for the scalar - $\rootkerr$ scattering, to leading order in coupling.
\subsection{Orbital Angular Impulse}

The leading order orbital angular impulse in the KMOC formalism is given by
\begin{align}
	\Delta L_i^{\mu \nu} =\frac{\hbar ^2}{4}\int \hat{d}^4 \bar{q}_1 \hat{d}^4 \bar{q}_2 \hat{\delta}(p_1\cdot \bar{q} _1)\hat{\delta}(p_2\cdot \bar{q}_2) e^{-i(b\cdot \bar{q}_2)} \bigg[\bigg(\tilde{p}_i \wedge \frac{\partial}{\partial \tilde{p}_i }\bigg)^{\mu\nu}+\bigg(p_i \wedge \frac{\partial}{\partial p_i }\bigg)^{\mu\nu} \bigg]\cr 
	\hat{\delta}^{(4)} (\bar{q}_1 + \bar{q}_2)\ \mathcal{A}_{4}(p_{1},p_{2}\rightarrow \tilde{p}_1,\tilde{p}_2) \,, \label{ang_imp_KMOC}
\end{align}
where $p_i$'s are initial momenta and we denote the final momenta as  $\Tilde{p}_i =p_i +\hbar \bar{q}_i$. Here $\mathcal{A}_{4}(p_{1},p_{2}\rightarrow \tilde{p}_1,\tilde{p}_2)$ is the four-point scalar$-\rootkerr$ scattering amplitude given in \eqref{linearimpulseamplitude}. Since we express the amplitude as a function of $(p_i,\bq_i)$, we shall treat them as independent variables and consider the transformation $(p_i, \Tilde{p}_i) \rightarrow (p'_i,q'_i)$ and then set $p'_i=p_i$ to obtain the correct differential operator for the angular impulse. With
\begin{align}
	p'_i =p_i\,, \qquad q'_i =\tilde{p}_i -p_i\,,
\end{align}
we obtain the differential operators in new variables by treating $p_i =p_i(p'_i,q'_i)$ and $\tilde{p}_i =\tilde{p}_i(p'_i,q'_i)$. We find 
\begin{align}
	\partial _{p_i}^\mu :=\frac{\partial}{\partial p_{i\mu} } =\partial _{p'_i}^\mu -\partial _{q'_i}^\mu \,, \qquad  \partial _{\tilde{p}_i}^\mu :=\frac{\partial}{\partial \tilde{p}_{i\mu} } =\partial _{q'_i}^\mu \,.
\end{align}
Using these transformations, we write the orbital angular impulse where we treat $p'_i=p_i$ and $q'_i$ as independent variables\footnote{We have for convenience replaced $p'_{i} = p_{i}$ and $q'_{i} = q_{i}$ from now on.}
\begin{align}\label{64}
	\Delta L_i^{\mu \nu} =\frac{\hbar ^2}{4}\int \hat{d}^4 \bar{q}_1 \hat{d}^4 \bar{q}_2 \hat{\delta}(p_1\cdot \bq _1)\hat{\delta}(p_2\cdot \bq _2) e^{-ib\cdot \bq_2} \left[(p_i \wedge \partial_{p_i} )^{\mu\nu} +(\bq_i \wedge \partial _{\bq_i})^{\mu \nu} \right]\cr 
	\left\lbrace \hat{\delta}^{(4)} (\bar{q}_1 + \bar{q}_2)\mathcal{A}_{4}(p_{1},p_{2}\rightarrow p_1+\hbar\bq _1,p_2+\hbar\bq _2) \right\rbrace \,. 
\end{align}

\subsubsection{Orbital angular impulse of the scalar particle}

For the scalar particle, we do integration by parts on the second term in the first line in eq.\eqref{64} and integrate over $\bq_{2}$ to obtain
\begin{align}
	\Delta L_1^{\mu \nu} =\Delta L_{1,I}^{\mu \nu} +\Delta L_{1,II}^{\mu \nu} \,,\label{ang_imp_KMOC1}
\end{align}  
with 
\begin{align} \label{scalar_orbimp}
	\Delta L_{1,I}^{\mu \nu}&= \frac{\hbar ^2}{4}  \int \hat{d}^4 \bar{q} e^{i\bar{q}\cdot b}\hat{\delta}(p_1 \cdot \bar{q})\hat{\delta}(p_2 \cdot \bar{q})\left(p_1 \wedge \frac{\partial}{\partial p_1} \right)^{\mu \nu} \mathcal{A}_4 (p_1,p_2 \rightarrow p_1+\hbar\bar{q},p_2-\hbar\bar{q})\cr
	\Delta L_{1,II}^{\mu \nu } &=-\frac{\hbar ^2}{4}\int \hat{d}^4 \bar{q} e^{i\bar{q}\cdot b}\ \hat{\delta}^{'}(p_1\cdot \bar{q})\ \hat{\delta}(p_2 \cdot \bar{q}) (\bar{q}\wedge p_1)^{\mu \nu}\ \mathcal{A}_4 (p_1,p_2 \rightarrow p_1+\hbar\bar{q},p_2-\hbar\bar{q})\,.
\end{align}
where ${\cal A}_{4}$ is given in eq. \eqref{linearimpulseamplitude}. As 
\begin{align}
	\frac{\partial}{\partial p_{j}^{\mu}}\, p_{i}^{\alpha}\, =\, \delta_{i}^{j}\, \delta_{\mu}^{\alpha}, \qquad \partial_{p_{1}^{\mu}}\, a_{2}^{\alpha} = 0 \,, \nonumber
\end{align} 
we will suppress the explicit dependence of $a_{2}^\mu$ on $p_{2}^\mu$ as in the radiation kernel derivation.

It is straightforward to evaluate the expression in $\Delta L_{1,I}^{\mu \nu}$.  Since $p_1$ and $\bq$ are independent, we get
\begin{align}
	\Scale[0.97]{\Delta L_{1,I}^{\mu \nu} 
		=Q_1Q_2 \int \hat{d}^4 \bar{q} e^{i\bar{q}\cdot b}\hat{\delta}(p_1 \cdot \bar{q})\hat{\delta}(p_2 \cdot \bar{q}) \frac{1}{\bq ^2} \left[(p_1\wedge p_2)^{\mu \nu} \cosh (a_2\cdot \bq)+i\frac{\sinh (a_2\cdot \bar{q})}{(a_2\cdot \bar{q})} p_1^{[\mu} \epsilon^{\nu]}(p_2,a_2,\bq)    \right]}\,. \label{orbitalscalar}
\end{align}
Evaluation of $\Delta L_{1,II}^{\mu \nu}$ on the other hand needs a more careful analysis since it involves derivative of the on-shell delta function.  In order to simplify this,  we shall decompose the momentum $\bq ^\mu$ along $p_{1,2}$ and in the transverse direction 
\begin{align}
	\bq^\mu = \alpha_1 p_1^\mu + \alpha_2 p_2^\mu + \bq_\perp^\mu \,, \qquad  p_i \cdot \bq_\perp =0\,,\label{massless_decomp}
\end{align}
where the coefficients are given by
\begin{align}
	\alpha_1 = \frac{1}{\mathcal{D}} [(p_1 \cdot p_2)x_2 - m_2^2 x_1]\,,\quad  \alpha_2 = \frac{1}{\mathcal{D}} [(p_1 \cdot p_2) x_1 - m_1^2 x_2] \,,
\end{align}
with $x_{1,2}:=(p_{1,2} \cdot \bq) $.  Due to this change of variables,  the measure transforms as follows
\begin{align}
	\hat{d}^4 \bq =\frac{1}{\sqrt{\mathcal{D}}} \hat{d}^2 \bq_\perp dx_1 dx_2\,,  \qquad \mathcal{D}=(p_1\cdot p_2)^2-m_1^2m_2^2\,.
\end{align}
In terms of $x_{1,2}$ and $\bq_\perp$ variables, we rewrite
\begin{align}
	\Delta L_{1,II}^{\mu \nu } =-\frac{Q_1Q_2}{\sqrt{\mathcal{D}}} \int \hat{d}^2 \bq _\perp dx_1   \hat{\delta}'(x_1) \frac{e^{i\bq _\perp \cdot b} }{\bq ^2} (\bar{q}\wedge p_1)^{\mu \nu} \Big\lbrace (p_1\cdot p_2) \cosh (a_2\cdot \bar{q})\cr
	+i\frac{\sinh (a_2\cdot \bar{q})}{(a_2\cdot \bar{q})} \epsilon (p_1,p_2,a_2,\bq) \Big\rbrace \,,
\end{align}
where we have done the $x_2$ integral and used $b\cdot p_{1,2}=0$.  Next, we perform an integration by parts in $x_1$.  At this stage, we note that from eq. \eqref{massless_decomp} we can write
\begin{align}
	q^2 = \alpha_1^2 p_1^2 + \alpha_2^2 p_2^2 + q_\perp^2 + 2\alpha_1 \alpha_2 (p_1 \cdot p_2)
\end{align}
Therefore any first order derivative of $\frac{1}{\bq ^2}$ w.r.t $x_1$ or $x_2$ vanishes due to the on-shell delta function constraints: $x_1=x_2=0$.  Using $\partial _{x_1}(a_2\cdot \bq)=-\frac{m_2^2}{\mathcal{D}}(a_2\cdot p_1)$, we get 
\begin{align}
	\Scale[0.97]{\Delta L_{1,II}^{\mu \nu } =\frac{Q_1Q_2}{\sqrt{\mathcal{D}}}\int \hat{d}^2 \bq _\perp  e^{i\bq _\perp \cdot b}\frac{1}{\bq_\perp ^2} \Big[ \frac{p_1\cdot p_2}{\mathcal{D}} (p_2\wedge p_1)^{\mu \nu} \left\lbrace (p_1\cdot p_2) \cosh (a_2\cdot \bq_\perp)+i\frac{\sinh (a_2\cdot \bq_\perp)}{(a_2\cdot \bq_\perp)} \epsilon (p_1,p_2,a_2,\bq_\perp) \right\rbrace} \cr
	\qquad \Scale[0.97]{+\frac{m_2^2}{\mathcal{D}} (a_2\cdot p_1)(p_1  \wedge \bq_\perp )^{\mu \nu} \left\lbrace (p_1\cdot p_2) \sinh (a_2\cdot \bq_\perp)+i\mathcal{Y} \epsilon (p_1,p_2,a_2,\bq_\perp) \right\rbrace \Big]\,},
\end{align}
where $\mathcal{Y} =\left[ \frac{\cosh{( a_2 \cdot \bq_\perp)}}{(a_2 \cdot \bq_\perp)}  - \frac{\sinh{( a_2 \cdot \bq_\perp)}}{(a_2 \cdot \bq_\perp)^2}\right] $.
Summing the two expressions $\Delta L_{1,I}^{\mu \nu}$ and $\Delta L_{1,II}^{\mu \nu}$, we obtain the orbital angular impulse of the scalar particle 
\begin{align}\label{orb_imp_1_kmo}
	\Delta L_1^{\mu \nu}= Q_1Q_2 & \int  \hat{d}^4 \bq \hat{\delta}(p_1\cdot \bq) \hat{\delta}(p_2\cdot \bq) \frac{e^{i\bq \cdot b}}{\bq ^2}\Big[  (p_1\wedge p_2)^{\mu \nu} \cosh (a_2\cdot \bq )\left(1-\frac{(p_1\cdot p_2)^2}{\mathcal{D}} \right)\cr
	&+i\frac{\sinh (a_2\cdot \bq)}{(a_2\cdot \bq)}\left( p_1^{[\mu} \epsilon^{\nu]} (p_2,a_2,\bq) +\frac{p_1\cdot p_2}{\mathcal{D}}\epsilon (p_1,p_2,a_2,\bq) (p_2\wedge p_1)^{\mu \nu}\right)\cr
	&+\frac{m_2^2}{\mathcal{D}} (a_2\cdot p_1)(p_1  \wedge \bq)^{\mu \nu} \left\lbrace (p_1\cdot p_2) \sinh (a_2\cdot \bq)+i\mathcal{Y} \epsilon (p_1,p_2,a_2,\bq) \right\rbrace \Big]\,.
\end{align}
In Appendix \ref{sec:orb_imp_scalar} we have verified the result in classical theory.

We use the results of the integrals from appendix \ref{spinallorderintegrals} and obtain the orbital angular impulse as follows
\begin{align}\label{orb_scal_int}
	\Delta L_1^{\mu \nu} &=  \frac{Q_1Q_2}{2\pi\sqrt{\mathcal{D}}}~ \text{Re}~ \Bigg[\frac{1}{\gamma^2 \beta^2} (p_2 \wedge p_1)^{\mu\nu} \Big\{\Big( 1 +\sum_{n=1}\frac{(-a_2 \cdot i\partial_b)^{2n}}{(2n)!}\Big) \log |\mu_1 b| \Big\}\cr
	&+ \Big(p_1^{[\mu} \epsilon^{\nu]\sigma \alpha \beta} p_{2\alpha} a_{2\beta} + \frac{p_1\cdot p_2}{\mathcal{D}}\epsilon ^{\rho \sigma \alpha \beta}p_{1\rho}p_{2\alpha}a_{2\beta}  (p_2\wedge p_1)^{\mu \nu} \Big) \mathcal{X}_\sigma \cr
	&+\frac{m_2^2}{\mathcal{D}} (a_2\cdot p_1) \Big \{ i (p_1  \wedge (b+i\Pi a_2))^{\mu \nu}   \frac{(p_1\cdot p_2)}{(b+i\Pi a_2)^2} + \epsilon ^{\rho \sigma \alpha \beta}p_{1\rho}p_{2\alpha}a_{2\beta} \Big(p_1 \wedge \frac{\partial}{\partial a_2} \Big)^{\mu\nu}  \mathcal{X}_\sigma \Big\} \Bigg] \,. \cr
\end{align}
Here we have defined
\begin{align}
	\mathcal{X}_\sigma := \frac{b_\sigma}{b^2 + i (b\cdot a_2) } + i \frac{(\Pi a_2)_\sigma}{(\Pi a_2)^2} \Big[\frac{\Pi a_2}{\Pi a_2 - i b} + \log \Big|\frac{b}{b + i \Pi a_2}\Big|\Big] \,,
\end{align}
where $\Pi^{\nu} {}_\rho$ is the projector into the plane orthogonal to both $u_1$ and $u_2$ \cite{Maybee:2019jus},
\begin{align}
	\Pi^{\nu} {}_\rho = \delta^{\nu} {}_\rho + \frac{1}{\gamma^2\beta^2} [u_1^\nu(u_{1\rho} - \gamma u_{2\rho}) + u_2^\nu(u_{2\rho} - \gamma u_{1\rho})] \,,
\end{align} 
with $\Pi a_2 = \sqrt{\Pi a_2 \cdot \Pi a_2}$ and $b = \sqrt{-b^2}$.
In eq.\eqref{orb_scal_int}, $\mu_1$ is the infrared (IR) cut-off. Note that the spin dependent terms in first line are not IR divergent as it involves derivative over $b$ and the scalar term contributes to the electromagnetic scoot.

The angular impulse for the scalar particle to linear order in spin written in terms of $S_2^{\perp\mu\nu}$ is
\begin{align} \label{scalarlinear}
	\Delta L_{1}^{\mu\nu} =  \frac{Q_1 Q_2}{2\pi \sqrt{\mathcal{D}}} \Big[\frac{1}{\beta ^2 \gamma ^2}(p_2 \wedge p_1)^{\mu\nu}\log 
	|\mu_1 b| +\frac{1}{b^2} \Big(p_1^{[\mu} S_2^{\perp\nu] \rho}b_{\rho} + (p_2 \wedge p_1)^{\mu\nu} \frac{(p_1 \cdot p_2)}{\mathcal{D}}  S_2^{\perp\rho\sigma}p_{1\rho}b_{\sigma}\Big)\Big] \cr
	+ \mathcal{O}(S_2^{\perp2}) \,.
\end{align}


\subsubsection{Orbital angular impulse of the $\rootkerr$ particle } \label{rootkerr_angimp}

The integral expression for the leading order orbital angular impulse of $\rootkerr$ particle can be written as,  
\begin{align}
	\Scale[0.97]{\Delta L_2^{\mu \nu} =\frac{\hbar ^2}{4}\int \hat{d}^4 \bar{q} e^{i\bar{q}\cdot b}\hat{\delta}(p_1 \cdot \bar{q})\hat{\delta}(p_2 \cdot \bar{q}) \left[\left(p_2 \wedge \frac{\partial}{\partial p_2} \right)^{\mu \nu} - i (\bar{q}\wedge b)^{\mu \nu} \right]\mathcal{A}_4 (p_1,p_2 \rightarrow p_1+\hbar\bar{q},p_2-\hbar\bar{q})}\cr
	\Scale[0.97]{- \frac{\hbar ^2}{4}\int \hat{d}^4 \bar{q} e^{i\bar{q}\cdot b}\hat{\delta}(p_1 \cdot \bar{q}) \hat{\delta}^{'}(p_2\cdot \bar{q})\ (\bar{q}\wedge p_2)^{\mu \nu}\ \mathcal{A}_4 (p_1,p_2 \rightarrow p_1+\hbar\bar{q},p_2-\hbar\bar{q})}.
\end{align}
We use the formula for linear impulse to rewrite the above integral as follows
\begin{align}
	\Delta L_2^{\mu \nu} = -(b \wedge \Delta p_2)^{\mu \nu} + \Delta L_{2,I}^{\mu \nu}
	+\Delta L_{2,II}^{\mu \nu} \,,\label{ang_imp_KMOC2}
\end{align}
where 
\begin{align}\label{620}
	\Delta L_{2,I}^{\mu \nu} &=\frac{\hbar ^2}{4}  \int \hat{d}^4 \bar{q} e^{i\bar{q}\cdot b}\hat{\delta}(p_1 \cdot \bar{q})\hat{\delta}(p_2 \cdot \bar{q})\left(p_2 \wedge \frac{\partial}{\partial p_2} \right)^{\mu \nu} \mathcal{A}_4 (p_1,p_2 \rightarrow p_1+\hbar\bar{q},p_2-\hbar\bar{q})\,,\cr
	\Delta L_{2,II}^{\mu \nu }&=-\frac{\hbar ^2}{4}\int \hat{d}^4 \bar{q} e^{i\bar{q}\cdot b}\ \hat{\delta}(p_1 \cdot \bar{q})\ \hat{\delta}^{'}(p_2\cdot \bar{q})\ (\bar{q}\wedge p_2)^{\mu \nu}\ \mathcal{A}_4 (p_1,p_2 \rightarrow p_1+\hbar\bar{q},p_2-\hbar\bar{q})\,.\cr
	\Delta p^{\mu}_2 &= \frac{\hbar^{2}}{4}   \int \hat{d}^{4}\bar{q}  \hat{\delta}(p_{1}\cdot \bar{q} )\hat{\delta}(p_{2}\cdot \bar{q})\ e^{i\bar{q}\cdot b}\ (-i\bar{q}^{\mu})\ \mathcal{A}_{4}(p_{1},p_{2}\rightarrow p_{1}+\hbar\bar{q},p_{2}-\hbar\bar{q}) 
\end{align}
The evaluation of $\Delta L_{2, I}^{\mu\nu}$ is rather subtle as for any function 
$f(a_{2},\bq)$ we obtain terms involving $\frac{\partial}{\partial p_{2}^{\mu}}\, f\vert_{a_{2} (S_{2}^{\perp},\, p_{2})}$. 
\begin{align}
	\frac{\partial}{\partial p_{2}^{\mu}}\, f(a_{2}, \bq)\, =\, -\frac{1}{2m_2^2}\frac{\partial f(a_{2},\bq)}{\partial a_{2}^{\alpha}}\epsilon_{\mu}{}^{\alpha\rho\sigma}S_{2\rho\sigma}^{\perp}  \,,
\end{align}
where we have used
\begin{align}
	\frac{\partial a_2^\alpha}{\partial p_2^\mu} = \frac{1}{2m_2^2}\epsilon^{\alpha\beta\rho\sigma}S_{2\rho\sigma}^{\perp}\delta_{\mu\beta} \,, \nonumber
\end{align} using the dual relation \eqref{11}.
The derivation of $\Delta L_{2,I}^{\mu\nu}$ (and hence $\Delta L_{2}^{\mu\nu}$)  (for $\vert a_{2} \vert\, \sim\, \vert b \vert$) will be pursued elsewhere \cite{Akhtar:2024rzp}. In this paper, we simply evaluate the orbital angular impulse to linear order in $S_{2}^{\perp\mu\nu}$.

Using the dual relation \eqref{11}, we obtain 
\begin{align}\label{646}
	\Delta L_{2}^{\mu \nu} &= Q_1Q_2 \int \hat{d}^4 \bq \hat{\delta}(\bar{q} \cdot p_1)\hat{\delta}(\bar{q} \cdot p_2)   e^{i\bq \cdot b}\frac{1}{\bq ^2} \Big[\frac{1}{\beta ^2 \gamma ^2}(p_1 \wedge p_2)^{\mu\nu}-(b\wedge \bq)^{\mu \nu}\ (S_2^{\perp\rho\sigma}p_{1\rho} \bq_{\sigma}) \cr
	&\hspace{6cm}+i(p_1 \wedge p_2)^{\mu\nu} \frac{(p_1 \cdot p_2)}{\mathcal{D}}  S_2^{\perp\rho\sigma}p_{1\rho}\bq_{\sigma} \Big]\cr
	&= \frac{Q_1 Q_2}{2\pi \sqrt{\mathcal{D}}} \Big[\frac{1}{\beta ^2 \gamma ^2}(p_1 \wedge p_2)^{\mu\nu}\log 
	|\mu_2 b| - \frac{1}{b^2} \Big(b^{[\mu} S_2^{\perp\nu] \rho}p_{1\rho} - (p_1 \wedge p_2)^{\mu\nu} \frac{(p_1 \cdot p_2)}{\mathcal{D}}  S_2^{\perp\rho\sigma}p_{1\rho}b_{\sigma}\Big)\Big] \,,
\end{align}
where $\mu_2$ is the IR cutoff.
This matches with the result obtained in classical theory given in Appendix \ref{sec:orb_imp_rootkerr_class}.

\subsection{Spin Angular Impulse }

The computation of the spin angular impulse $\Delta S_{2}^{\perp\mu\nu}$ is rather straightforward via the NJ algorithm. Using the inverse of the dual relation in eq.\eqref{11},
\begin{equation}
	S^{\perp\mu\nu} = \epsilon^{\mu\nu\rho\sigma}p_{\rho}a_{\sigma}
\end{equation}
we obtain \cite{Luna:2023uwd}
\begin{align}
	\Delta S_2^{\perp\mu\nu} = \epsilon^{\mu\nu\rho\sigma}\Delta p_{2\rho} a_{2\sigma} + \epsilon^{\mu\nu\rho\sigma}p_{2\rho} \Delta a_{2\sigma} \,,\label{spin_imp_dual}
\end{align}
where $\Delta a_{2}^{\mu}$ is known as the spin kick.  We note that although $S_{2}^{\perp\mu\nu}$ is the fundamental spin degree of freedom, the NJ algorithm lets us directly compute the spin kick which can then be used to deduce $\Delta S_{2}^{\perp\mu\nu}$.
Since the linear impulse doesn't receive any radiative contribution at leading order in the coupling, the expression for the linear impulse $\Delta p_2^\mu$ is exactly opposite to $\Delta p_1^\mu$, derived in section \ref{sec:lin_imp} and it is given by
\begin{align}
	\Delta p_2^{\mu}
	= - iQ_1Q_2  \int \hat{d}^4 \bar{q} \hat{\delta}(\bar{q} \cdot u_1)\hat{\delta}(\bar{q} \cdot u_2)\frac{e^{i \bar{q} \cdot b}}{\bar{q}^2} \Big[ \gamma \cosh{( a_2 \cdot \bar{q})}\bar{q}^{\mu} + i \sinh ( a_2 \cdot \bar{q}) \epsilon^\mu(\bar{q},u_1,u_2) \Big] \,,
\end{align}
where we rewrite the $\sinh (a_2 \cdot \bq)$ term using the identity
\begin{align}
	(a_2\cdot \bq) \epsilon_{\nu}(u_1,u_2,\bq) = \bq_\nu \epsilon(u_1,u_2,a_2,\bq) \,.
\end{align}
We study the leading order spin kick using the following formula \cite{Maybee:2019jus}
\begin{align} \label{spinkickallorder}
	\Delta a_2^\mu &= \Big \llangle \frac{i\hbar^2}{4} \int \hat{d}^4 \bq \hat{\delta}(p_1 \cdot \bq)\hat{\delta}(p_2 \cdot \bq) e^{i\bq \cdot b} \Big\{ \left[a_2^\mu (p_2) , \mathcal{A}_4 \right] + \frac{\hbar}{m_2}(a_2\cdot \bq)u_2^\mu \mathcal{A}_4  \Big\}\Big \rrangle\,.
\end{align} 
The commutator in the first term is defined in the SU(2) little group space. The SU(2) indices are left implicit under the double angle bracket notation, explained in section \ref{KMOCsetup}. Note that, the formula \eqref{spinkickallorder} appears to be non-uniform in the order of $\hbar$, however, the commutator term also includes an additional factor of $\hbar$ and it is given by 
\begin{align}
	[a_{2,IK}^\mu , a_{2,KJ}^\nu] = \frac{i\hbar}{m_2} \epsilon^{\mu\nu \rho \sigma} u_{2\rho}a_{2\sigma ,IJ}\,, \label{a-commutator}
\end{align}
where we display the SU(2) indices $(I,J,K)$. 
From hereafter, we shall drop the SU(2) indices and the double angle bracket notation altogether. Using \eqref{a-commutator}, we get
\begin{align}
	[a_2^\mu, \cosh(a_2 \cdot \bq)] = \frac{i\hbar}{m_2} \sinh{(a_2 \cdot \bq)} \epsilon^\mu (\bq,u_2,a_2) \,.
\end{align}
Next we consider the following commutator
\begin{align}
	\Big[a_2^\mu, \frac{\sinh(a_2 \cdot \bq)}{(a_2 \cdot \bq)}\epsilon(u_1,u_2,a_2,\bq)\Big] &= \frac{i\hbar}{m_2} \mathcal{Y} \epsilon^{\mu}(\bq,u_2,a_2)\epsilon(u_1,u_2,a_2,\bq) \cr
	&+ \frac{i\hbar}{m_2} \frac{\sinh{( a_2 \cdot \bq)}}{(a_2 \cdot \bq)}\epsilon^{\mu\nu}(u_2,a_2)\epsilon_{\nu}(u_1,u_2,\bq) \,. 
\end{align}
where we defined $\mathcal{Y}:=\Big(\frac{\cosh{( a_2 \cdot \bq)}}{(a_2 \cdot \bq)}  - \frac{\sinh{( a_2 \cdot \bq)}}{(a_2 \cdot \bq)^2}\Big)$.  Using eq. \ref{identities} it can be shown that,
\begin{align}
	(a_2\cdot \bq) \epsilon_{\nu}(u_1,u_2,\bq) = \bq_\nu \epsilon(u_1,u_2,a_2,\bq) \,, 
\end{align}
to get
\begin{align}
	\Big[a_2^\mu, \frac{\sinh(a_2 \cdot \bq)}{(a_2 \cdot \bq)}\epsilon(u_1,u_2,a_2,\bq)\Big] &= \frac{i \hbar}{m_2} \frac{\cosh{(a_2 \cdot \bq)}}{(a_2 \cdot \bq)} \epsilon^\mu (\bq,u_2,a_2) \epsilon(u_1,u_2,a_2, \bq)\,. 
\end{align}
Next, we use the identity
\begin{align}
	\epsilon^\mu (\bq,u_2,a_2) \epsilon(u_1,u_2,a_2, \bq) =(a_2\cdot \bq) \left[(a_2\cdot \bq)(u_1^\mu -\gamma u_2^\mu)-\bq^\mu (a_2\cdot u_1) \right]\,,
\end{align}
where we have set $(a_2\cdot u_2)=0=(u_{1,2}\cdot \bq)$ and $(u_1\cdot u_2)=\gamma$, to finally get
\begin{align}
	\Big[a_2^\mu, \frac{\sinh(a_2 \cdot \bq)}{(a_2 \cdot \bq)}\epsilon(u_1,u_2,a_2,\bq)\Big] = \frac{i \hbar}{m_2} \cosh{(a_2 \cdot \bq)} [-\gamma u_2^\mu (a_2 \cdot \bq)+ u_1^\mu (a_2 \cdot \bq) -\bq^\mu (a_2 \cdot u_1)] \,.
\end{align}
Substituting various commutator expressions, we obtain the leading order spin kick as
\begin{align} \label{spinkickall}
	\Delta a_2^\mu =  \frac{i Q_1 Q_2}{m_2}\int \hat{d}^4 \bq \frac{1}{\bq^2} \hat{\delta}(u_1 \cdot \bq)\hat{\delta}(u_2 \cdot \bq) e^{i\bq \cdot b} \Big\{\cosh{(a_2 \cdot \bq)} [u_1^\mu (a_2 \cdot \bq) - \bq^\mu (a_2 \cdot u_1)] \cr 
	- i\sinh{(a_2 \cdot \bq)} \epsilon^\mu (u_1,a_2,\bq) \Big\} \,,
\end{align}
where we make use of the identity
\begin{align}
	u_2^\mu \epsilon(u_1,u_2,a_2,\bq) =\gamma \epsilon^{\mu}(u_2,a_2,\bq) -\epsilon^{\mu} (u_1,a_2,\bq) \,.
\end{align}
This matches with the spin kick obtained in \cite{Guevara:2020xjx} using classical equations of motion. 
Plugging these expressions into \eqref{spin_imp_dual},
we obtain the following expression for spin angular impulse 

\begin{align} \label{spinangularallorder}
	\Delta S_2^{\perp\mu\nu} &= - Q_1Q_2  \int \hat{d}^4 \bar{q} \hat{\delta}(\bar{q} \cdot u_1)\hat{\delta}(\bar{q} \cdot u_2)\frac{e^{i \bar{q} \cdot b}}{\bar{q}^2} \Big[ \cosh{( a_2 \cdot \bar{q})} \Big\{ i u_1^{[\mu} \epsilon^{\nu]}(\bq,u_2,a_2) \cr
	&- i \bq^{[\mu} \epsilon^{\nu]}(u_{1},u_2,a_2) \Big\} + \sinh ( a_2 \cdot \bar{q}) \Big\{u_2^{[\mu}u_1^{\nu]}(a_{2}\cdot \bq) - u_2^{[\mu}\bq^{\nu]}(a_{2}\cdot u_{1})  + \gamma a_{2}^{[\mu}\bq^{\nu]}  \Big\} \Big] \,. \cr
\end{align}
Using the integral results, we get 
\begin{align}\label{spin_result}
	\Delta S^{\perp\mu\nu}_{2} &= \frac{Q_1 Q_2}{2\pi\gamma\beta}\text{Re}~ \Big[ \frac{1}{(b + i\Pi a_2)^2}\Big\{b^{[\mu} \epsilon^{\nu]} (u_1,u_2,a_2) -u_1^{[\mu}\epsilon^{\nu]}(u_2,a_2,b) +u_2^{[\mu}u_1^{\nu]}a_2^2 \cr & \hspace{-0.5cm}+\frac{1}{\gamma^2 \beta^2}u_2^{[\mu} a_2^{\nu]}(a_2 \cdot u_1)
	+\frac{(a_2 \cdot u_1)}{\gamma \beta^2} a_2^{[\mu} u_1^{\nu]} +i \Big(a_2^{[\mu} \epsilon^{\nu]} (u_1,u_2,a_2) - \frac{(a_2 \cdot u_1)}{\gamma \beta^2} u_2^{[\mu}\epsilon^{\nu]} (u_1,u_2,a_2)\cr
	& -u_2^{[\mu}u_1^{\nu]} (a_2 \cdot b)+ u_2^{[\mu}b^{\nu]}(a_2 \cdot u_1) + \gamma a_2^{[\mu}b^{\nu]}\Big)\Big\}\Big]  \,.
\end{align}
It can be verified that the RHS of eq.(\ref{spin_result}) is $\Delta S_{2}^{\perp \mu\nu}$ as it satisfies the SSC constraint to leading order in the coupling. 
\begin{align}\label{ssc_chk}
	\Delta S_2^{\perp\mu \nu}p_{2\nu}+S_2^{\perp\mu \nu}\Delta p_{2\nu}=0\,.
\end{align}
At linear order in $S_{2}^{\perp\mu\nu}$, the spin angular impulse can be evaluated. 
\begin{align} \label{spinimpulselinear}
	\Delta S_{2}^{\perp\mu\nu} &= -iQ_1Q_2 \int \hat{d}^4 \bq \hat{\delta}(p_1\cdot \bq)\hat{\delta}(p_2\cdot \bq) e^{i \bq \cdot b} \frac{1}{\bq^2} \Big[p_1^{[\mu}S_2^{\perp\nu]\sigma}\bq_{\sigma} - \bq^{[\mu}S_2^{\perp\nu]\sigma}p_{1\sigma} \Big] + \mathcal{O}(S_2^{\perp 2})
\end{align}
\begin{equation}\label{637}
	\Delta S_{2}^{\perp\mu\nu} = \frac{Q_1 Q_2}{2\pi \sqrt{\mathcal{D}}\ b^2}\big(b^{[\mu}S^{\perp\nu]\alpha}p_{1\alpha} - p_{1}^{[\mu}S^{\perp\nu]\alpha}b_{\alpha}\big) .
\end{equation}
This expression is in agreement with the result of \cite{Bern:2023ity}, which was derived using the classical equations of motion.

We now have all the expressions to compute the total angular impulse, $\Delta J^{\mu\nu}$ in eq.\eqref{totalangcons} for the the scalar-$\rootkerr$ scattering to linear order in spin. This is given by the sum of eqs.\eqref{scalarlinear},\eqref{646} and \eqref{spinimpulselinear}. We obtain the following result
\begin{align}\label{631}
	\Delta J^{\mu\nu} &= \frac{Q_1 Q_2}{2\pi \sqrt{\mathcal{D}}}\frac{1}{\beta ^2 \gamma ^2}(p_1 \wedge p_2)^{\mu\nu}\log \Big|\frac{\mu_2}{\mu_1} \Big| + \delta^{\mu\nu}_{\textrm{scalar-scoot}} = 0 \,,
\end{align}
where 
\begin{align}
	\delta^{\mu\nu}_{\textrm{scalar-scoot}} = -\frac{Q_1 Q_2}{2\pi \sqrt{\mathcal{D}}}\frac{1}{\beta ^2 \gamma ^2}(p_1 \wedge p_2)^{\mu\nu}\log \Big|\frac{\tau_1}{\tau_2} \Big| \,.
\end{align}
The IR cutoffs are related to the proper times of the two particles via $\frac{\mu_2}{\mu_1} = \frac{\tau_1}{\tau_2}$ \cite{DiVecchia:2022piu,Bhardwaj:2022hip}.
Hence, the total angular momentum for scalar - $\rootkerr$ scattering is conserved, to linear order in spin. The study of conservation of angular momentum to all orders in spin is under investigation \cite{Akhtar:2024rzp}.
\section{Discussion}\label{discussion}

Building upon the synthesis of the Newman-Janis (NJ) algorithm with the KMOC formalism in \cite{Arkani-Hamed:2019ymq}, in this work, we have used the NJ algorithm to compute classical observables beyond the linear impulse for electromagnetic scattering involving $\rootkerr$ particles. As is well known, the real power of the NJ algorithm lies in all orders in spin results for classical observables. We hope that by combining the on-shell methods along with the NJ algorithm one can build ``loop integrands'' associated with the scattering of $\rootkerr$ particles such that our analysis can be extended beyond leading order in the coupling. For some early attempts in this direction, see \cite{Menezes:2022tcs}. 

Our main idea is to simply re-interpret the three-point coupling involving $\rootkerr$ as a ``spin dressing'' of the photon polarisation data while computing higher point amplitudes.  Our work shows that the resulting ``spin-dressed''  photon propagator is particularly useful for constructing the five-point amplitude where the photon is emitted from the external scalar state.  We used this five-point amplitude to derive the radiation emitted by the scalar to all orders in the spin and found perfect agreement with its calculation using equations of motion. 

We then used the four-point amplitude computed using the spin-dressed photon propagator to compute the angular impulse of the scalar as well as $\rootkerr$ particles.  Here we encountered a subtlety. We used the spin tensor $S_{2}^{\mu\nu}$ instead of the spin-vector $a_2^{\mu}$ as the fundamental spin degree of freedom. With this choice, we found that the result for the angular impulse of the $\rootkerr$ particle is consistent with the angular momentum conservation. 
As explained in Section \ref{sec7} the reason for this choice becomes evident if one recalls that the calculation of the angular impulse via the KMOC formalism involves the expectation value of differential operators acting on amplitudes. Choosing the spin to be parametrized by either $a_2$ or $S_2$  leads to different results as a consequence.  
To leading order in $S_{2}$ (valid as long as $\vert a_{2}\vert \ll \vert b\vert$) we have checked that our results are consistent with \cite{Bern:2023ity}. 

Using the NJ algorithm, we also calculated the leading-order orbital angular impulse of a scalar particle to all order in the spin of the $\rootkerr$ particle. In addition we gave a closed form expression for the total angular impulse of the $\rootkerr$ particle to leading order in spin. An all-order-in spin evaluation of the total angular momentum of the $\rootkerr$ particle is beyond the scope of this paper and will be pursued in \cite{Akhtar:2024rzp}.\\
Our broader goal is to compute classical gravitational observables for the Kerr black hole. In this context, the double copy will be an important tool \cite{Bern:2019prr}. For conservative observables (at leading order in coupling), the double copy of the three-point amplitudes is the only ingredient that is needed. It has been observed in \cite{Arkani-Hamed:2017jhn} that the three-point amplitude for a massive spin - $S$ minimally coupled to gravity can be obtained by simply squaring the `$x$' factors in eq.\eqref{35}. Using this in \cite{Arkani-Hamed:2019ymq}, the 1PM linear impulse for the scalar-Kerr system was obtained. It will be interesting to use this double copy to study the angular impulse for the same system and check the conservation of angular momentum. However, to study radiation from gravitational scattering involving Kerr black holes, we need the double copy of the non-abelian counterpart of the $\rootkerr$ solution. Since we do not have a consistent bare Lagrangian for the latter we leave the question of computing gravitational radiation for Kerr black holes from amplitudes, to future work. In the meantime, it might be instructive to study gravitational radiation from Kerr black holes directly using the equations of motion derived in \cite{Scheopner:2023rzp}. However, recently the waveform for the scattering of a Kerr$_{1}$ - Kerr$_{2}$ black hole system has been computed in both the perturbative spin parameter to $\mathcal{O}(a^{4})$, the highest order to which the answer is unambiguous\cite{DeAngelis:2023lvf}. The case of a Schwarzschild black hole scattering with a Kerr black hole has also been studied in \cite{Brandhuber:2023hhl} and the waveform has been computed for the Schwarzschild black hole to all orders in spin and for the Kerr black hole to $\mathcal{O}(a^{4})$. The former case is the gravitational analogue of the radiative gauge field in eq.\eqref{radker_scala_kmo_exacta1}. Additionally, the waveform for this system has also been computed to NLO in \cite{Bohnenblust:2023qmy} to linear order in spin. Finally, it is also important to note that in \cite{Aoude:2023dui}, the waveform for the Kerr$_{1}$ - Kerr$_{2}$ system has been computed to all orders in spin for both the black holes. These result will be interesting to compare with, once we have a result from our NJ perspective.

\acknowledgments
We are grateful to Alok Laddha for suggesting the problem and numerous insightful discussions. We thank him for carefully going through the draft and providing valuable suggestions and comments on it. We thank Sujay Ashok for constant encouragement and valuable comments on the draft. We thank all the participants of `Amplitudes @ Chennai' workshop at the Chennai Mathematical Institute for interesting discussions. SA would like to thank the organizers of the `Indian Strings Meeting 2023' at the Indian Institute of Technology Bombay for an opportunity to present a part of this work as a poster, and the participants for their feedback. We also thank Yilber Fabian Bautista and Rafael A. Porto for their feedback and discussion. AM thanks Aninda Sinha for support and acknowledges financial support from DST through the SERB core grant CRG/2021/000873.

\begin{appendix}
\section{Conventions}\label{convention}
Throughout the paper, we will use the metric signature as $(+,-,-,-)$, unless otherwise stated. So, the on-shell condition is $p^2=m^2$. Since the impact parameter is  spacelike we have $-b^2 > 0$. We will use $\epsilon_{0123} = +1$. The rescaled delta functions appearing in section \ref{KMOCsetup} are defined as
\begin{equation}
	\hat{\delta}(p\cdot q) := 2\pi\delta(p\cdot q),\ \ \ \ \ \hat{\delta}^{(4)}(p + q) := (2\pi)^{4}\delta^{(4)}(p + q).
\end{equation}
where $p^{\mu}$ and $q^{\mu}$ are generic four vectors. We also absorb the $2\pi$ factor in the measure $d^4q$ and define the rescaled the rescaled measure as
\begin{equation}
	\hat{d}^{4}q := \frac{d^{4}q}{(2\pi)^{4}}.
\end{equation}
The anti-symmetric bracket in all the expressions are defined as
\begin{align}
	(A \wedge B)^{\mu\nu} = A^{[\mu} B^{\nu]} = A^{\mu} B^{\nu} - B^{\mu} A^{\nu} \,.
\end{align} 
We use the following compact notations in the main text for convenience
\begin{align}
	\epsilon^{\mu\nu} (A,B) &= \epsilon^{\mu\nu\alpha\beta}A_\alpha B_\beta \cr 			\epsilon^\mu (A,B,C) &= \epsilon^{\mu\nu\alpha\beta}A_\nu B_\alpha C_\beta \cr
	\epsilon(A,B,C,D) &= \epsilon^{\mu\nu\alpha\beta}A_\mu B_\nu C_\alpha D_\beta \,,
\end{align}
where $(A^\mu ,B^\mu ,C^\mu ,D^\mu )$ are generic 4-vectors. The following identity is used throughout this paper.
\begin{align}\label{identities}  
	A^{[\mu} \epsilon ^{\nu]} (B,C,D) &=-(A\cdot B)\epsilon ^{\mu \nu}(C,D)-(A\cdot C)\epsilon ^{\mu \nu}(D,B)-(A.D)\epsilon ^{\mu \nu}(B,C)\,.
\end{align}

\section{Orbital angular impulse in KMOC formalism}\label{angularmomentumderiv}

In this appendix, we derive the expression for the change in the orbital angular momentum in the KMOC formalism, given in eq.\eqref{ang_momentum_exp}. The change in the angular momentum of the particles can be defined analogous to the linear impulse
\begin{equation}
\Delta L_{i}^{\mu\nu} = i\bra{\Psi} [\Ll_{i}^{\mu\nu},\mathbb{T}]\ket{\Psi} + \bra{\Psi}\mathbb{T}^{\dag}[\Ll_{i}^{\mu\nu},\mathbb{T}]\ket{\Psi}.
\end{equation}
We are interested only in the leading order term and hence, we concentrate only on the first term. Plugging in the expression for the initial state in eq.\eqref{ini}, we get
\begin{equation}\label{B6}
\begin{split}
\Delta L_{i}^{\mu\nu}  = i\int \prod_{i=1}^{2}\ d\Phi(r_{i})d\Phi(p_{i})\ &e^{-ir_{2}\cdot b/\hbar}e^{ip_{2}\cdot b/\hbar}\ \phi^{*}(r_{i})\phi(p_{i})\\
&(\bra{\vec{r}_{1},\vec{r}_{2}}\Ll_{i}^{\mu\nu}\mathbb{T}\ket{\vec{p}_{1},\vec{p}_{2}} - \bra{\vec{r}_{1},\vec{r}_{2}}\mathbb{T}\Ll_{i}^{\mu\nu}\ket{\vec{p}_{1},\vec{p}_{2}})\, .
\end{split}
\end{equation}
Here we have suppressed the little group indices $a_i$. Now, we use
\begin{equation}\label{B7}
\bra{\vec{r}_{1},\vec{r}_{2}}\mathbb{T}\Ll_{i}^{\mu\nu}\ket{\vec{p}_{1},\vec{p}_{2}} \coloneqq i\hbar \bigg(p_{i}\wedge \frac{\partial}{\partial p_{i}}\bigg)^{\mu\nu}\bra{\vec{r}_{1},\vec{r}_{2}}\mathbb{T}\ket{\vec{p}_{1},\vec{p}_{2}},
\end{equation}
to write the first term in terms of the differential operator, we use the hermiticity of the orbital angular momentum operator i.e
\begin{align}
\bra{\vec{r}_{1},\vec{r}_{2}}\Ll_{i}^{\mu\nu}\mathbb{T}\ket{\vec{p}_{1},\vec{p}_{2}} & = \big(\bra{\vec{p}_{1},\vec{p}_{2}}\mathbb{T}^{\dag}\Ll_{i}^{\mu\nu}\ket{\vec{r}_{1},\vec{r}_{2}}\big)^{\dag}\\
& = -i\hbar\bigg(r_{i}\wedge \frac{\partial}{\partial r_{i}}\bigg)^{\mu\nu}(\bra{\vec{p}_{1},\vec{p}_{2}}\mathbb{T}^{\dag}\ket{\vec{r}_{1},\vec{r}_{2}})^{\dag}\\
& = -i\hbar\bigg(r_{i}\wedge \frac{\partial}{\partial r_{i}}\bigg)^{\mu\nu}\bra{\vec{r}_{1},\vec{r}_{2}}\mathbb{T}\ket{\vec{p}_{1},\vec{p}_{2}}
\end{align}
Plugging the last equation and eq.\eqref{B7} into eq.\eqref{B6}, we get
\begin{equation}
\begin{split}
\Delta L_{i}^{\mu\nu}  = \hbar\int \prod_{i=1}^{2}\ d\Phi(r_{i})d\Phi(p_{i})\ e^{-ir_{2}\cdot b/\hbar}&e^{ip_{2}\cdot b/\hbar}\ \phi^{*}(r_{i})\phi(p_{i})\\
&\bigg[\bigg(p_{i}\wedge \frac{\partial}{\partial p_{i}}\bigg)^{\mu\nu} + \bigg(r_{i}\wedge \frac{\partial}{\partial r_{i}}\bigg)^{\mu\nu}\bigg]\bra{\vec{r}_{1},\vec{r}_{2}}\mathbb{T}\ket{\vec{p}_{1},\vec{p}_{2}}
\end{split}
\end{equation}
By relabelling from $r_{i} = p_{i} + q_{i}$, we get
\begin{equation}
\begin{split}
\braket{\Delta L_{i}^{\mu\nu}} = \hbar\int \prod_{i=1}^{2}\ \hat{d}^{4}q_{i}\ & \hat{\delta}(2p_{i}\cdot q_{i} + q_{i}^{2})\ e^{-iq_{2}\cdot b/\hbar}\\
& \bigg(\bigg(p_{i}\wedge \frac{\partial}{\partial p_{i}}\bigg)^{\mu\nu}+\bigg((p_{i}+q_{i})\wedge \frac{\partial}{\partial (p_{i}+q_{i})}\bigg)^{\mu\nu}\bigg) 
				A_{4}(p_{1},p_{2}\rightarrow p_{1}+q_{1},p_{2}+q_{2})\ \, 
\end{split}
\end{equation}  
where $\braket{\Delta L_{i}^{\mu\nu}}$ denotes the integration over the wave functions. This is the same expression in eq.\eqref{ang_momentum_exp}, where we have substituted the defnition of the 4-point scattering amplitude
\begin{equation}
A_{4}(p_{1},p_{2}\rightarrow p_{1}+q_{1},p_{2}+q_{2}):=\bra{\vec{p}_{1}+\vec{q}_{1},\vec{p}_{2}+\vec{q}_{2}}\mathbb{T}\ket{\vec{p}_{1},\vec{p}_{2}} \,.
\end{equation}

\section{Classical calculations}
In this appendix, we present the classical calculations of angular impulse and the radiation kernel for scalar in scalar-$\rootkerr$ scattering to leading order in coupling.

\subsection{Equations of motion}
We present all the equations of motion that will be used to derive the physical observables discussed in the main text.
The equation of motion for the scalar particle in scalar-$\rootkerr$ scattering \cite{Guevara:2020xjx} is 
\begin{align}
	\frac{dp_1^\mu}{d\tau}=Q_1 \text{Re}F_{2,+}^{\mu \nu}(x+ia_2)u_{1\nu}\,,
\end{align}
where $F_{+}^{\mu \nu}(x+ia_2)$ is the self-dual part of the electromagnetic field strength of the $\rootkerr$ particle.  The self dual and anti-self dual field strengths are defined w.r.t to Minkowski metric as follows
\begin{align}
	F_{\pm}^{\mu \nu}(x+ia_2)=F^{\mu \nu}\pm \frac{i}{2}\epsilon ^{\mu \nu \rho \sigma}F_{\rho \sigma} \,.\label{sef_dual_field}
\end{align}
Note that, the self dual and anti-self dual fields are related to each other via complex conjugation \footnote{We assume that (for all $c_n$ be real)
	\begin{align}
		F_{\mu \nu}(x\pm i a)=\sum _{n}c_n(x \pm i a)^n \Rightarrow [F_{\mu \nu}(x\pm i a)]^\dagger =F_{\mu \nu}(x\mp i a) \,.
\end{align}}
\begin{align}
	F^-_{\mu \nu}(x-i a) =\left[F^+_{\mu \nu}(x+i a)\right]^\dagger\,.
\end{align}
Therefore we can rewrite the real part of the self dual field strength as
\begin{align}
	2 \text{Re}F_{2,+}^{\mu \nu}(x_1 +ia_2)= F_+^{\mu \nu}(x_1+ia_2)+ F_-^{\mu \nu}(x_1-ia_2)
\end{align}
We then use the definition \eqref{sef_dual_field} to express the equation of motion as follows
\begin{align}
	\frac{dp_1^\mu}{d\tau} =Q_1 \left[\cos (a_2\cdot \partial)F^{\mu \nu}(x)-\frac{1}{2}\epsilon ^{\mu \nu \rho \sigma} \sin (a_2\cdot \partial)F_{\rho \sigma}(x)\right]u_{1\nu}\,.\label{lorentzeqnexacta}
\end{align}
Note that, the field strength appearing here is due to a charged scalar particle!
We now use the following expression for the field strength in momentum space 
\begin{align}
	F_2^{\mu \nu} (\bq) &=(iQ_2) e^{-i\bq\cdot b} \hat{\delta}(u_2\cdot \bq)\frac{1}{\bq^2} (\bq^\mu u_2^\nu -\bq^\nu u_2^\mu)\,,
\end{align}
to substitute in \eqref{lorentzeqnexacta} to obtain an expression for $\dot{p}_1^\mu$ to all order in spin $a_2^\mu$
\begin{align}
	\dot{p}_1^\mu (\tau)=i Q_1 Q_2 \int \hat{d}^4 \bq \hat{\delta}(u_2\cdot \bq) e^{-i\bq\cdot x(\tau)}\frac{e^{-i\bq\cdot b}}{\bq^2} \Big[ \cosh(a_2\cdot \bq) \lbrace \gamma \bq^\mu -u_2^\mu (u_1\cdot \bq)\rbrace \cr
	+i\sinh(a_2\cdot \bq)\epsilon ^{\mu \nu \rho \sigma}u_{1\nu}  \bq_{\rho}u_{2\sigma } \Big]\,.\label{p1dot_class}
\end{align}

\subsection{Orbital angular impulse for the scalar particle}  \label{sec:orb_imp_scalar}
Classically, the orbital angular impulse is defined as 
\begin{align}
	\frac{d L^{\mu \nu}}{d\tau}=(x\wedge \dot{p})^{\mu \nu}\,,
\end{align}
where $p^\mu =m \dot{x}^\mu $. Now using the parametrization  of the classical trajectory of particle 1 $ x_1^\mu (\tau)=u^{\mu}_1 \tau \,,$
the orbital angular momentum impulse to leading order in coupling is
\begin{align}
	\frac{d L_1^{\mu \nu}}{d\tau}=\tau (u_1 \wedge \dot{p}_1)^{\mu \nu} \,. \label{ang_mom_class_change}
\end{align}
We use the following identity from appendix \ref{identities}:
\begin{align}
	(a_2 \cdot \bq)\epsilon^{\mu}(p_1,p_2,\bq) = \bq^{\mu}\epsilon(p_1,p_2,a_2,\bq) - (p_1 \cdot \bq)\epsilon^{\mu}(a_2,\bq,p_2) \,,
\end{align} 
to rewrite \eqref{p1dot_class} as follows
\begin{align}
	\dot{p}_1^\mu (\tau)  =\frac{i Q_1 Q_2}{m_1} \int \hat{d}^4 \bq \hat{\delta}(p_2\cdot \bq)& \frac{e^{+i\bq\cdot b}}{\bq^2}e^{-i(\bq\cdot p_1)\frac{\tau}{m_1}} \Big[\cosh (a_2\cdot \bq) \lbrace  (p_1\cdot p_2) \bq^\mu - (p_1\cdot \bq)p_2^\mu \rbrace \cr
	&  \hspace{-1cm}+i \frac{\sinh(a_2\cdot \bq)}{a_2\cdot \bq}\lbrace (\bq\cdot p_1)\epsilon ^\mu (a_2,p_2,\bq)+\bq^\mu \epsilon (a_2,\bq,p_1,p_2) \rbrace   \Big]\,,
\end{align}
to derive the LO expression for the orbital angular impulse for the scalar particle
\begin{align}
	\Delta L_1^{\mu \nu} &=\frac{i Q_1 Q_2}{m_1} \int d\tau \hat{d}^4\bq \hat{\delta}(p_2\cdot \bq)\frac{e^{+i\bq\cdot b}}{\bq^2} \left( i \partial _{p_1\cdot \bq} e^{-i(\bq\cdot p_1)\frac{\tau}{m_1}}\right) \Big[\cosh (a_2\cdot \bq) \lbrace  (p_1\cdot p_2) (p_1 \wedge \bq)^{\mu \nu} \cr
	&- (p_1\cdot \bq)(p_1 \wedge p_2)^{\mu\nu} \rbrace  +i \frac{\sinh(a_2\cdot \bq)}{a_2\cdot \bq}\lbrace (\bq\cdot p_1)p_1^{[\mu}\epsilon ^{\nu ]} (a_2,p_2,\bq)+ (p_1 \wedge \bq)^{\mu \nu} \epsilon (a_2,\bq,p_1,p_2) \rbrace   \Big]\,.
\end{align}
Here we need the derivative w.r.t $u_1\cdot \bq$ due to the factor of $\tau$ in \eqref{ang_mom_class_change}.  Now in the classical computation, we shall replace
\begin{align}
	\bq^\mu =\alpha _1p_1^\mu +\alpha _2p_2^\mu +\bq_\perp ^\mu \,,\label{decomp_class_mass}
\end{align}
where the coefficients are given  in equation \eqref{massless_decomp}.  Again, we can do the $x_2 =p_2\cdot \bq$ integral in the above integral and write
\begin{align}
	\Delta L_1^{\mu \nu} &=-\frac{ Q_1 Q_2}{m_1\sqrt{\mathcal{D}}} \int d\tau d^2 \bq_\perp dx_1 \frac{e^{+i\bq_\perp \cdot b}}{\bq^2} \left( \partial _{x_1} e^{-i \frac{x_1 \tau}{m_1}}\right) \Big[\cosh (a_2\cdot \bq) \lbrace (p_1\cdot p_2) (p_1 \wedge \bq)^{\mu \nu} \cr
	&- (p_1\cdot \bq)(p_1 \wedge p_2)^{\mu\nu} \rbrace   +i \frac{\sinh(a_2\cdot \bq)}{a_2\cdot \bq} \lbrace x_1 p_1^{[\mu}\epsilon ^{\nu ]} (a_2,p_2,\bq)+ (p_1 \wedge \bq)^{\mu \nu} \epsilon (a_2,\bq,p_1,p_2) \rbrace    \Big]\, ,\cr
\end{align}
where $x_{1} = p_{1}\cdot \bq$. Integrating by parts in $x_1$ variable and then completing the $\tau$ integral, we obtain \begin{align}
	\Delta L_1^{\mu \nu} =\frac{ Q_1 Q_2}{\sqrt{\mathcal{D}}} & \int  d^2 \bq_\perp dx_1 \hat{\delta}(x_1) \frac{e^{i\bq_\perp \cdot b}}{\bq_\perp ^2} \Big[\frac{1}{\beta ^2 \gamma ^2} \cosh (a_2\cdot \bq)(p_1\wedge p_2)^{\mu \nu}\cr
	& -i \frac{\sinh (a_2\cdot \bq)}{a_2\cdot \bq} \left\lbrace  p_1^{[\mu}S_2^{\nu ]\sigma } \bq_\sigma +\frac{p_1\cdot p_2}{\mathcal{D}}  (p_2 \wedge p_1)^{\mu \nu}S_2^{\perp\rho\sigma}p_{1\rho}\bq_{\perp\sigma} \right\rbrace  \cr
	&-\frac{m_2^2 (a_2\cdot p_1)}{\mathcal{D}} \left[ \sinh (a_2\cdot \bq)(p_1\cdot p_2)+i \mathcal{Y} S_2^{\perp\rho\sigma}p_{1\rho}\bq_{\perp\sigma}\right] (p_1\wedge \bq_\perp)^{\mu \nu} \Big]
\end{align}
To linear order in spin, we find
\begin{align}
	\Delta L_{1,\mathcal{O}(S_2)}^{\mu \nu} =-\frac{i Q_1 Q_2}{\sqrt{\mathcal{D}}} \int  d^2 \bq_\perp  \frac{e^{+i\bq_\perp \cdot b}}{\bq_\perp ^2} \left\lbrace  p_1^{[\mu}S_2^{\nu ]\sigma } \bq_\sigma +\frac{p_1\cdot p_2}{\mathcal{D}}  (p_2 \wedge p_1)^{\mu \nu}S_2^{\perp\rho\sigma}p_{1\rho}\bq_{\perp\sigma} \right\rbrace
\end{align}

\subsection{Angular impulse for $\rootkerr$ particle} \label{sec:orb_imp_rootkerr_class}
The orbital angular impulse for the $\rootkerr$ particle is given by
\begin{align}
	\Delta L_2^{\mu \nu} =(b\wedge \Delta p_2)^{\mu \nu} +\int d\tau \tau (u_2\wedge \dot{p}_2)^{\mu \nu}=(b\wedge \Delta p_2)^{\mu \nu}+I^{\mu \nu} \,.  \label{ang_imp_class_2}
\end{align}
Using the equation of motion for the $\rootkerr$ particle to linear order in spin
\begin{align}
	\dot{p}_{2}^\rho  &=Q_1 F_1^{\rho \sigma}p_{2\sigma}- \frac{Q_1}{2}S_2^{\perp\mu c} \partial ^\rho F_{1,\mu c}\,,
\end{align}
we obtain the following integral expression for $\dot{p}_{2\rho}$ 
\begin{align}\label{p2dot_class}
	\dot{p}_{2\rho} = -\frac{iQ_1 Q_2}{m_2} \int \hat{d}^4 \bq e^{i \bq \cdot b} e^{i (p_2 \cdot \bq)\frac{\tau}{m_2}}\frac{\hat{\delta}(p_1 \cdot \bq)}{\bq^2} \Big[(p_1 \cdot p_2)\bq_\rho - (p_2 \cdot \bq)p_{1\rho} + i \bq_\rho S_2^{\perp\rho\sigma}p_{1\rho}\bq_{\sigma} \Big]\,.
\end{align}
We use this expression in \eqref{ang_imp_class_2} to obtain 
\begin{align}
	I^{\mu\nu} = -\frac{Q_1 Q_2}{m_2} \int \hat{d}^4 \bq d\tau e^{i \bq \cdot b} \frac{\partial}{\partial (p_2 \cdot \bq)} \Big(e^{i (p_2 \cdot \bq)\frac{\tau}{m_2}} \Big) \frac{\hat{\delta}(p_1 \cdot \bq)}{\bq^2} \Big[(p_1 \cdot p_2)(p_2 \wedge \bq)^{\mu\nu} - (p_2 \cdot \bq)(p_2 \wedge p_1)^{\mu\nu} \cr
	+ i (p_2 \wedge \bq)^{\mu\nu} S_2^{\perp\rho\sigma}p_{1\rho}\bq_{\sigma} \Big] \,.
\end{align}
Again, in order to evaluate this integral we use the decomposition in \eqref{decomp_class_mass} and following similar steps as we did in evaluating $\Delta L_1^{\mu \nu}$.  Finally, we expand $\Delta p_2^\mu$ to $\mathcal{O}(S_2)$ and get the orbital angular impulse as 
\begin{align} \label{kerrlinear}
	\Delta L_2^{\mu\nu}  = \frac{Q_1 Q_2}{\sqrt{\mathcal{D}}} \int \hat{d}^2 \bq_\perp e^{i \bq_\perp \cdot b} \frac{1}{\bq_\perp^2} \Big[\frac{1}{\beta ^2 \gamma ^2} + i \frac{(p_1 \cdot p_2)}{\mathcal{D}} S_2^{\perp\rho\sigma}p_{1\rho}\bq_{\perp\sigma}\Big](p_2 \wedge p_1)^{\mu\nu} \cr
	-i\frac{Q_1 Q_2}{\sqrt{\mathcal{D}}} \int \hat{d}^2 \bq_\perp e^{i \bq_\perp \cdot b} \frac{1}{\bq_\perp^2} (b \wedge \bq_\perp)^{\mu\nu} \Big[(p_1 \cdot p_2) + i S_2^{\perp\rho\sigma}p_{1\rho}\bq_{\perp\sigma} \Big] \,.
\end{align}

\subsection{Radiation from the scalar particle} \label{sec:radiation_scalar}
The classical current from particle 1 in momentum space is given by
\begin{align}
	J_1^\mu (x)=Q_{1}\int d\tau_1  e^{i\bar{k}\cdot x_1(\tau_1)} \frac{i}{\bar{k}\cdot p_1}\left[ \dot{p}_1^\mu -\frac{\bar{k}\cdot \dot{p}_1}{\bar{k} \cdot p_1}p_1^\mu \right]\,.
\end{align}
Using the expression in \eqref{p1dot_class}, we obtain the current to all order in spin
\begin{align}
	J_1^\mu (\bar{k},a_2)&=Q_{1}^{2}Q_{2} \int \hat{d}^4 \bq \hat{\delta}(u_1\cdot \bq-u_{1}\cdot \bar{k}) \hat{\delta}(u_2\cdot \bq) \frac{e^{-i\bq\cdot b}}{\bq^2}\frac{1}{\bar{k}\cdot p_1}\cr 
	& \times\Big[  \cosh(a_2\cdot \bq) \lbrace \gamma \bq^\mu -u_2^\mu (u_1\cdot \bq)\rbrace+i\sinh(a_2\cdot \bq)\epsilon ^{\mu \nu \rho \sigma}u_{1\nu}  \bq_{\rho}u_{2\sigma }  \cr
	& -\frac{p_1^\mu}{\bar{k}\cdot p_1}\big\lbrace \cosh (a_2\cdot \bq)\left( \gamma (\bar{k}\cdot \bq)- (\bar{k}\cdot u_2) (u_1\cdot \bq)\right)  + i\sinh (a_2\cdot \bq) \epsilon (\bar{k},u_1,\bq,u_2)  \big \rbrace \Big]\,.\cr  \label{radker_scalar_exacta}
\end{align}

We now take a soft limit of the current and show that, on comparing with the classical sub-leading soft factor, it reproduces the angular impulse for the scalar particle.
We use the following identity from Appendix \ref{identities}:
\begin{align}
	(a_2 \cdot \bq)\epsilon^{\mu}(u_1,u_2,\bq) = \bq^{\mu}\epsilon(u_1,u_2,a_2,\bq) - (u_1 \cdot \bar{k})\epsilon^{\mu}(a_2,\bq,u_2) \,.
\end{align}
The classical current is then rewritten as
\begin{align}
	J^{\mu}_1 (\bar{k},a_2) &=Q_{1}^{2}Q_{2} \int \hat{d}^4 \bar{q} \hat{\delta}[u_1\cdot (\bar{q}-\bar{k})] \hat{\delta}(u_2\cdot \bar{q}) \frac{e^{-i\bar{q}\cdot b}}{\bar{q}^2}\frac{1}{\bar{k}\cdot p_1}\cr 
	& \times\Big[  \cosh(a_2\cdot \bar{q}) \Big\{ \gamma \bar{q}^\mu -u_2^\mu (u_1\cdot \bar{k}) - \frac{p_1^\mu}{p_1 \cdot \bar{k}} \Big(\gamma (\bar{k} \cdot \bq) - (\bar{k} \cdot u_2)(\bar{k} \cdot u_1)  \Big)\Big\} \cr
	&-i\frac{\sinh(a_2\cdot \bar{q})}{a_2 \cdot \bq} \Big\{\bq^\mu \epsilon(u_1,u_2,a_2,\bq) - (u_1 \cdot \bar{k}) \epsilon^{\mu}(a_2,\bar{q},u_2)  \cr
	&-\frac{p_1^\mu}{\bar{k}\cdot p_1}\Big( (\bq \cdot \bar{k})\epsilon(u_1,u_2,a_2,\bq) - (u_1 \cdot \bar{k})\epsilon(\bar{k},a_2,\bq,u_2) \Big) \Big\}\Big]\,.\cr
\end{align}

Next, we do a soft expansion in $\bar{k}$ where $k^{\mu} = \omega(1,\hat{n})$. We expand the delta function as follows
\begin{align}
	\hat{\delta}(p_1 \cdot \bq -p_1\cdot \bar{k})=\hat{\delta}(p_1 \cdot \bq)-(p_1\cdot \bar{k})\hat{\delta}'(p_1 \cdot \bq)+\mathcal{O}(\bar{k}^2)\,,
\end{align}
and write the sub-leading soft terms of the classical current as follows
\begin{align}
	J^{\mu}_1 (\bar{k},a_2)|_{\mathcal{O}(\omega^{0})} &=Q_{1}^{2}Q_{2} \int \hat{d}^4 \bar{q} \hat{\delta}(p_1\cdot \bar{q}) \hat{\delta}(p_2\cdot \bar{q}) \frac{e^{-i\bar{q}\cdot b}}{\bar{q}^2}  \Big[ \cosh(a_2\cdot \bar{q}) \Big\{-p_2^\mu +p_1^\mu \frac{(\bar{k}\cdot p_2)}{(\bar{k}\cdot p_1)}\Big\} \cr
	&\qquad \qquad -i\frac{\sinh(a_2\cdot \bar{q})}{a_2 \cdot q} \Big\{-\epsilon^{\mu}(a_2,\bar{q},p_2) + \frac{p_1^\mu}{(\bar{k} \cdot p_1)} \epsilon(\bar{k},a_2,\bq,p_2)  \Big\} \Big]\cr
	&-Q_{1}^{2}Q_{2} \int \hat{d}^4 \bar{q} \hat{\delta}^\prime(p_1\cdot \bar{q}) \hat{\delta}(p_2\cdot \bar{q}) \frac{e^{-i\bar{q}\cdot b}}{\bar{q}^2} \Big[\cosh(a_2\cdot \bar{q}) (p_1 \cdot p_2)\Big\{  \bar{q}^\mu - \frac{p_1^\mu}{p_1 \cdot \bar{k}} (\bar{k} \cdot \bq) \Big\} \cr
	&\qquad \qquad -i\frac{\sinh(a_2\cdot \bar{q})}{a_2 \cdot \bq} \Big\{\bq^\mu \epsilon(p_1,p_2,a_2,\bq) - \frac{p_1^\mu}{\bar{k}\cdot p_1}(\bq \cdot \bar{k})\epsilon(p_1,p_2,a_2,\bq)  \Big\}\Big]\,. \cr
\end{align}

Comparing with the classical sub-leading soft factor \cite{Manu:2020zxl},
\begin{align}
	S_1^{(1)\mu} = Q_1 \Big[ \frac{\Delta J_1^{\mu\nu} \bar{k}_\nu}{(p_1 \cdot \bar{k})} - \frac{(\Delta p_1 \cdot \bar{k})}{(p_1 \cdot \bar{k})^2}J_{-1}^{\mu\nu} \bar{k}_\nu\Big] \,,
\end{align}
where $\Delta p_1^\mu$ is the LO linear impulse of the scalar particle, $\Delta J_1^{\mu\nu}$ is the LO total angular impulse of the scalar particle and $J_{-1}^{\mu\nu}$ is the initial angular momentum tensor, we obtain
\begin{align}
	\Delta J_1^{\mu\nu} &= Q_1 Q_2 \int  \hat{d}^4 \bq e^{-i \bq \cdot b} \hat{\delta}(p_1 \cdot \bq)\hat{\delta}(p_2 \cdot \bq) \frac{1}{\bq^2} \Big[(p_1 \wedge p_2)^{\mu\nu} \cosh (a_2 \cdot \bq) -i\frac{ \sinh{(a_2 \cdot \bq)}}{(a_2 \cdot \bq)}  p_1^{[\mu}S_2^{\perp\nu]\sigma}\bq_\sigma \Big] \cr
	&-Q_1 Q_2 \int  \hat{d}^4 \bq e^{-i \bq \cdot b} \hat{\delta}^\prime(p_1 \cdot \bq)\hat{\delta}(p_2 \cdot \bq) \frac{1}{\bq^2} (\bq \wedge p_1)^{\mu\nu} \Big[\cosh{ (a_2 \cdot \bq)}(p_1 \cdot p_2) \cr
	&\qquad \qquad \qquad-i \frac{\sinh{( a_2 \cdot \bq)}}{(a_2 \cdot \bq)} S_2^{\perp\rho\sigma}p_{1\rho}\bq_{\sigma} \Big] \,. 
\end{align}
Replacing $\bq \rightarrow -\bq$, we get
\begin{align}
	\Delta J_1^{\mu\nu} &= Q_1 Q_2 \int  \hat{d}^4 \bq e^{i \bq \cdot b} \hat{\delta}(p_1 \cdot \bq)\hat{\delta}(p_2 \cdot \bq) \frac{1}{\bq^2} \Big[(p_1 \wedge p_2)^{\mu\nu} \cosh (a_2 \cdot \bq) +i\frac{ \sinh{(a_2 \cdot \bq)}}{(a_2 \cdot \bq)}  p_1^{[\mu}S_2^{\perp\nu]\sigma}\bq_\sigma \Big] \cr
	&-Q_1 Q_2 \int  \hat{d}^4 \bq e^{i \bq \cdot b} \hat{\delta}^\prime(p_1 \cdot \bq)\hat{\delta}(p_2 \cdot \bq) \frac{1}{\bq^2} (\bq \wedge p_1)^{\mu\nu} \Big[\cosh{ (a_2 \cdot \bq)}(p_1 \cdot p_2) \cr
	&\qquad \qquad \qquad +i \frac{\sinh{( a_2 \cdot \bq)}}{(a_2 \cdot \bq)} S_2^{\perp\rho\sigma}p_{1\rho}\bq_{\sigma} \Big] \,, 
\end{align}
which matches with the computation of angular impulse of the scalar particle of eq. \eqref{scalar_orbimp}. Here $J_{-1}^{\mu\nu} = (b_1 \wedge p_1)^{\mu\nu} =0$ in our setup as $b_1 =0, b_2 =b$.

\section{Evaluation of Integrals}\label{spinallorderintegrals}
In this appendix, we perform the integrals required to calculate the angular impulse for the scalar$-\rootkerr$ scattering. 
To obtain an expression for $\Delta L_1^{\mu \nu}$ from eq.\eqref{orb_imp_1_kmo}, we need to evaluate a series of integrals which are discussed below. 
\begin{itemize}
	\item We start with the following integral
	\begin{align}
		I_1 &=  \int \hat{d}^4 \bar{q} \hat{\delta}(\bar{q} \cdot u_1)\hat{\delta}(\bar{q} \cdot u_2)\frac{e^{i \bar{q} \cdot b}}{\bar{q}^2} \cosh (a_2 \cdot q) \cr
		&=   \int \hat{d}^4 \bar{q} \hat{\delta}(\bar{q} \cdot u_1)\hat{\delta}(\bar{q} \cdot u_2)\frac{e^{i \bar{q} \cdot b}}{\bar{q}^2} \Big(1 +  \sum_{n=1}\frac{(a_2 \cdot q)^{2n}}{(2n)!} \Big) \cr
		&= \frac{1}{2\pi\gamma\beta} \Big[ 1 +\sum_{n=1}\frac{(-a_2 \cdot i\partial_b)^{2n}}{(2n)!}\Big] \log |\mu b| \,,
	\end{align}
	where $\mu$ is the infrared (IR) cutoff. Note that the spin dependent terms are not IR divergent as it involves derivative over $b$.
	\item Next, we consider
	\begin{align}
		I_{2,\sigma} &= i  \int \hat{d}^4 \bar{q} \hat{\delta}(\bar{q} \cdot u_1)\hat{\delta}(\bar{q} \cdot u_2)\frac{e^{i \bar{q} \cdot b}}{\bar{q}^2} \frac{\sinh{( a_2 \cdot q)}}{(a_2 \cdot q)}  q_\sigma 
		\cr
		&=  \frac{i}{2}  \int \hat{d}^4 \bar{q} \hat{\delta}(\bar{q} \cdot u_1)\hat{\delta}(\bar{q} \cdot u_2)\frac{e^{i \bar{q} \cdot b}}{\bar{q}^2} \int_0^1 d\lambda \Big( e^{\lambda(a_2 \cdot q)} + e^{-\lambda (a_2 \cdot q)} \Big)q_\sigma \cr
		&= ~ \text{Re}~ \int \hat{d}^4 \bar{q} \hat{\delta}(\bar{q} \cdot u_1)\hat{\delta}(\bar{q} \cdot u_2) \int_0^1 d\lambda  e^{i \bar{q} \cdot (b+i \lambda a_2)}\frac{i}{\bar{q}^2}q_\sigma \cr
		&= \frac{1}{2\pi\gamma\beta}~  \text{Re} \int_0^1 d\lambda \frac{(b+ i \lambda \Pi a_2)_\sigma}{b^2+2i \lambda (b\cdot \Pi a_2)-\lambda ^2 (\Pi a_2)^2}\,.
	\end{align}
	Now we do the $\lambda$ integral as follows
	\begin{align}
		& \int_0^1 d\lambda \frac{(b+ i \lambda \Pi a_2)_\sigma}{b^2+2i \lambda (b\cdot \Pi a_2)-\lambda ^2 (\Pi a_2)^2}\cr
		&=b_\sigma \int_0^1  \frac{d\lambda}{b^2+2i \lambda (b\cdot \Pi a_2)-\lambda ^2 (\Pi a_2)^2}+i (\Pi a_2)_\sigma \int_0^1  \frac{\lambda d\lambda}{b^2+2i \lambda (b\cdot \Pi a_2)-\lambda ^2 (\Pi a_2)^2} \cr
		&= \frac{b_\sigma}{(b^2 + i (b\cdot \Pi a_2) )} + i \frac{(\Pi a_2)_\sigma}{(\Pi a_2)^2} \Big(\frac{\Pi a_2}{\Pi a_2 - i b} + \log \Big|\frac{b}{b + i \Pi a_2}\Big| \Big) \,.
	\end{align}
	Therefore, we get 
	\begin{align}\label{int_I2}
		I_{2,\sigma} &= \frac{1}{2\pi\gamma\beta}~  \text{Re} \Big[ \frac{b_\sigma}{(b^2 + i (b\cdot \Pi a_2) )} + i \frac{(\Pi a_2)_\sigma}{(\Pi a_2)^2} \Big(\frac{\Pi a_2}{\Pi a_2 - i b} + \log \Big|\frac{b}{b + i \Pi a_2}\Big| \Big) \Big] \,.
	\end{align} 
	
	\item We consider the integral
	\begin{align}\label{int_I3}
		I_3^\alpha &= \int  \hat{d}^4 q e^{i q \cdot b} \hat{\delta}(u_1 \cdot q)\hat{\delta}(u_2 \cdot q) \frac{1}{q^2} \sinh (a_2 \cdot q) q^\alpha \cr
		&= \frac{1}{2} \int  \hat{d}^4 q e^{i q \cdot b} \hat{\delta}(u_1 \cdot q)\hat{\delta}(u_2 \cdot q) \frac{1}{q^2} q^\alpha \Big(e^{(a_2 \cdot q)} - e^{-(a_2 \cdot q)} \Big) \cr
		&= \frac{1}{2\pi\gamma\beta}~ \text{Re}~ \Big[ \frac{i(b+i\Pi a_2)^{\alpha}}{(b+i\Pi a_2)^2} \Big] \,.
	\end{align}
	
	\item Lastly, we evaluate the integral
	\begin{align}
		I_4^{\alpha\beta} &= i \int  \hat{d}^4 q e^{i q \cdot b} \hat{\delta}(u_1 \cdot q)\hat{\delta}(u_2 \cdot q) \frac{1}{q^2}  \mathcal{Y} q^\alpha q^\beta \cr
		&= i \int  \hat{d}^4 q e^{i q \cdot b} \hat{\delta}(u_1 \cdot q)\hat{\delta}(u_2 \cdot q) \frac{1}{q^2}  \Big(\frac{\cosh{( a_2 \cdot q)}}{(a_2 \cdot q)}  - \frac{\sinh{( a_2 \cdot q)}}{(a_2 \cdot q)^2} \Big) q^\alpha q^\beta \cr
		&= i \frac{\partial}{\partial a_{2\alpha}} \int  \hat{d}^4 q e^{i q \cdot b} \hat{\delta}(u_1 \cdot q)\hat{\delta}(u_2 \cdot q) \frac{1}{q^2} \frac{\sinh{( a_2 \cdot q)}}{(a_2 \cdot q)} q^\beta \,.
	\end{align}
	Using the result for $I_{2,\sigma}$ in \eqref{int_I2}, we find
	\begin{align}
		I_4^{\alpha\beta} &= \frac{1}{2\pi\gamma\beta}\, \frac{\partial}{\partial a_{2\alpha}} \text{Re} \Big[ \frac{b^\beta}{(b^2 + i (b\cdot \Pi a_2) )} + i \frac{(\Pi a_2)^\beta}{(\Pi a_2)^2} \Big(\frac{\Pi a_2}{\Pi a_2 - i b} + \log \Big|\frac{b}{b + i \Pi a_2}\Big| \Big) \Big] \,.
	\end{align}
\end{itemize}	
We use the results of these integrals to derive the expression for $\Delta L_1^{\mu \nu}$ in \eqref{orb_scal_int}.
For the spin angular impulse presented in eq.\eqref{spinangularallorder}, we need the integral in \eqref{int_I3} and evaluate the following integral
\begin{align}
	I_5^{\mu\nu} &=i  \int \hat{d}^4 \bar{q} \hat{\delta}(\bar{q} \cdot u_1)\hat{\delta}(\bar{q} \cdot u_2)\frac{e^{i \bar{q} \cdot b}}{\bar{q}^2} \Big[ q^{[\mu} S_2^{\nu] \rho}u_{1\rho} -u_1^{[\mu}S_2^{\nu]\sigma}q_\sigma \Big] \cosh (a_2 \cdot q) \cr
	&= ~ \text{Re}~ \int \hat{d}^4 \bar{q} \hat{\delta}(\bar{q} \cdot u_1)\hat{\delta}(\bar{q} \cdot u_2)  e^{i \bar{q} \cdot (b+i a_2)}\frac{i}{\bar{q}^2} \Big[ q^{[\mu} S_2^{\nu] \rho}u_{1\rho} -u_1^{[\mu}S_2^{\nu]\sigma}q_\sigma \Big] \cr
	&=\frac{1}{2\pi\gamma\beta}~ \text{Re}~  \Big[\frac{(b+i\Pi a_2)^{[\mu} S_2^{\nu] \rho}u_{1\rho}-u_1^{[\mu}S_2^{\nu]\sigma}(b+ i \Pi a_2)_\sigma}{(b+i\Pi a_2)^2} \Big] \,.
\end{align}
We use this and \eqref{int_I3} to obtain the expression for $\Delta S_2^{\mu \nu}$ in \eqref{spin_result}.

Note that in all of the above integrals, $\Pi^{\nu} {}_\rho$ is the projector into the plane orthogonal to both $u_1$ and $u_2$,
\begin{align}
	\Pi^{\nu} {}_\rho = \delta^{\nu} {}_\rho + \frac{1}{\gamma^2\beta^2} [u_1^\nu(u_{1\rho} - \gamma u_{2\rho}) + u_2^\nu(u_{2\rho} - \gamma u_{1\rho})] \,,
\end{align} 
with $\Pi a_2 = \sqrt{\Pi a_2 \cdot \Pi a_2}$ and $b = \sqrt{-b^2}$.

\end{appendix}

\bibliographystyle{JHEP}
\bibliography{BHscattering}

\providecommand{\href}[2]{#2}\begingroup\raggedright\begin{thebibliography}{100}

\bibitem{Arkani-Hamed:2019ymq}
N.~Arkani-Hamed, Y.-t.~Huang and D.~O'Connell, \emph{{Kerr black holes as
  elementary particles}},
  \href{https://doi.org/10.1007/JHEP01(2020)046}{\emph{JHEP} {\bfseries 01}
  (2020) 046} [\href{https://arxiv.org/abs/1906.10100}{{\ttfamily
  1906.10100}}].

\bibitem{Bern:2019nnu}
Z.~Bern, C.~Cheung, R.~Roiban, C.-H.~Shen, M.P.~Solon and M.~Zeng,
  \emph{{Scattering Amplitudes and the Conservative Hamiltonian for Binary
  Systems at Third Post-Minkowskian Order}},
  \href{https://doi.org/10.1103/PhysRevLett.122.201603}{\emph{Phys. Rev. Lett.}
  {\bfseries 122} (2019) 201603}
  [\href{https://arxiv.org/abs/1901.04424}{{\ttfamily 1901.04424}}].

\bibitem{Bern:2019crd}
Z.~Bern, C.~Cheung, R.~Roiban, C.-H.~Shen, M.P.~Solon and M.~Zeng, \emph{{Black
  Hole Binary Dynamics from the Double Copy and Effective Theory}},
  \href{https://doi.org/10.1007/JHEP10(2019)206}{\emph{JHEP} {\bfseries 10}
  (2019) 206} [\href{https://arxiv.org/abs/1908.01493}{{\ttfamily
  1908.01493}}].

\bibitem{Bern:2020gjj}
Z.~Bern, H.~Ita, J.~Parra-Martinez and M.S.~Ruf, \emph{{Universality in the
  classical limit of massless gravitational scattering}},
  \href{https://doi.org/10.1103/PhysRevLett.125.031601}{\emph{Phys. Rev. Lett.}
  {\bfseries 125} (2020) 031601}
  [\href{https://arxiv.org/abs/2002.02459}{{\ttfamily 2002.02459}}].

\bibitem{Bern:2020buy}
Z.~Bern, A.~Luna, R.~Roiban, C.-H.~Shen and M.~Zeng, \emph{{Spinning black hole
  binary dynamics, scattering amplitudes, and effective field theory}},
  \href{https://doi.org/10.1103/PhysRevD.104.065014}{\emph{Phys. Rev. D}
  {\bfseries 104} (2021) 065014}
  [\href{https://arxiv.org/abs/2005.03071}{{\ttfamily 2005.03071}}].

\bibitem{Bern:2020uwk}
Z.~Bern, J.~Parra-Martinez, R.~Roiban, E.~Sawyer and C.-H.~Shen, \emph{{Leading
  Nonlinear Tidal Effects and Scattering Amplitudes}},
  \href{https://doi.org/10.1007/JHEP05(2021)188}{\emph{JHEP} {\bfseries 05}
  (2021) 188} [\href{https://arxiv.org/abs/2010.08559}{{\ttfamily
  2010.08559}}].

\bibitem{Bern:2021dqo}
Z.~Bern, J.~Parra-Martinez, R.~Roiban, M.S.~Ruf, C.-H.~Shen, M.P.~Solon et~al.,
  \emph{{Scattering Amplitudes and Conservative Binary Dynamics at ${\cal
  O}(G^4)$}}, \href{https://doi.org/10.1103/PhysRevLett.126.171601}{\emph{Phys.
  Rev. Lett.} {\bfseries 126} (2021) 171601}
  [\href{https://arxiv.org/abs/2101.07254}{{\ttfamily 2101.07254}}].

\bibitem{Bern:2021yeh}
Z.~Bern, J.~Parra-Martinez, R.~Roiban, M.S.~Ruf, C.-H.~Shen, M.P.~Solon et~al.,
  \emph{{Scattering Amplitudes, the Tail Effect, and Conservative Binary
  Dynamics at O(G4)}},
  \href{https://doi.org/10.1103/PhysRevLett.128.161103}{\emph{Phys. Rev. Lett.}
  {\bfseries 128} (2022) 161103}
  [\href{https://arxiv.org/abs/2112.10750}{{\ttfamily 2112.10750}}].

\bibitem{Bern:2021xze}
Z.~Bern, J.P.~Gatica, E.~Herrmann, A.~Luna and M.~Zeng, \emph{{Scalar QED as a
  toy model for higher-order effects in classical gravitational scattering}},
  \href{https://doi.org/10.1007/JHEP08(2022)131}{\emph{JHEP} {\bfseries 08}
  (2022) 131} [\href{https://arxiv.org/abs/2112.12243}{{\ttfamily
  2112.12243}}].

\bibitem{Bern:2022kto}
Z.~Bern, D.~Kosmopoulos, A.~Luna, R.~Roiban and F.~Teng, \emph{{Binary Dynamics
  through the Fifth Power of Spin at O(G2)}},
  \href{https://doi.org/10.1103/PhysRevLett.130.201402}{\emph{Phys. Rev. Lett.}
  {\bfseries 130} (2023) 201402}
  [\href{https://arxiv.org/abs/2203.06202}{{\ttfamily 2203.06202}}].

\bibitem{Bern:2022jvn}
Z.~Bern, J.~Parra-Martinez, R.~Roiban, M.S.~Ruf, C.-H.~Shen, M.P.~Solon et~al.,
  \emph{{Scattering amplitudes and conservative dynamics at the fourth
  post-Minkowskian order}},
  \href{https://doi.org/10.22323/1.416.0051}{\emph{PoS} {\bfseries LL2022}
  (2022) 051}.

\bibitem{Barack:2023oqp}
L.~Barack et~al., \emph{{Comparison of post-Minkowskian and self-force
  expansions: Scattering in a scalar charge toy model}},
  \href{https://doi.org/10.1103/PhysRevD.108.024025}{\emph{Phys. Rev. D}
  {\bfseries 108} (2023) 024025}
  [\href{https://arxiv.org/abs/2304.09200}{{\ttfamily 2304.09200}}].

\bibitem{Bern:2023ccb}
Z.~Bern, E.~Herrmann, R.~Roiban, M.S.~Ruf, A.V.~Smirnov, V.A.~Smirnov et~al.,
  \emph{{Conservative binary dynamics at order $O(\alpha^5)$ in
  electrodynamics}},  \href{https://arxiv.org/abs/2305.08981}{{\ttfamily
  2305.08981}}.

\bibitem{Manohar:2022dea}
A.V.~Manohar, A.K.~Ridgway and C.-H.~Shen, \emph{{Radiated Angular Momentum and
  Dissipative Effects in Classical Scattering}},
  \href{https://doi.org/10.1103/PhysRevLett.129.121601}{\emph{Phys. Rev. Lett.}
  {\bfseries 129} (2022) 121601}
  [\href{https://arxiv.org/abs/2203.04283}{{\ttfamily 2203.04283}}].

\bibitem{KoemansCollado:2019ggb}
A.~Koemans~Collado, P.~Di~Vecchia and R.~Russo, \emph{{Revisiting the second
  post-Minkowskian eikonal and the dynamics of binary black holes}},
  \href{https://doi.org/10.1103/PhysRevD.100.066028}{\emph{Phys. Rev. D}
  {\bfseries 100} (2019) 066028}
  [\href{https://arxiv.org/abs/1904.02667}{{\ttfamily 1904.02667}}].

\bibitem{Cristofoli:2020uzm}
A.~Cristofoli, P.H.~Damgaard, P.~Di~Vecchia and C.~Heissenberg,
  \emph{{Second-order Post-Minkowskian scattering in arbitrary dimensions}},
  \href{https://doi.org/10.1007/JHEP07(2020)122}{\emph{JHEP} {\bfseries 07}
  (2020) 122} [\href{https://arxiv.org/abs/2003.10274}{{\ttfamily
  2003.10274}}].

\bibitem{DiVecchia:2021ndb}
P.~Di~Vecchia, C.~Heissenberg, R.~Russo and G.~Veneziano, \emph{{Radiation
  Reaction from Soft Theorems}},
  \href{https://doi.org/10.1016/j.physletb.2021.136379}{\emph{Phys. Lett. B}
  {\bfseries 818} (2021) 136379}
  [\href{https://arxiv.org/abs/2101.05772}{{\ttfamily 2101.05772}}].

\bibitem{DiVecchia:2021bdo}
P.~Di~Vecchia, C.~Heissenberg, R.~Russo and G.~Veneziano, \emph{{The eikonal
  approach to gravitational scattering and radiation at $ \mathcal{O}
  $(G$^{3}$)}}, \href{https://doi.org/10.1007/JHEP07(2021)169}{\emph{JHEP}
  {\bfseries 07} (2021) 169}
  [\href{https://arxiv.org/abs/2104.03256}{{\ttfamily 2104.03256}}].

\bibitem{Alessio:2022kwv}
F.~Alessio and P.~Di~Vecchia, \emph{{Radiation reaction for spinning black-hole
  scattering}},
  \href{https://doi.org/10.1016/j.physletb.2022.137258}{\emph{Phys. Lett. B}
  {\bfseries 832} (2022) 137258}
  [\href{https://arxiv.org/abs/2203.13272}{{\ttfamily 2203.13272}}].

\bibitem{DiVecchia:2022piu}
P.~Di~Vecchia, C.~Heissenberg, R.~Russo and G.~Veneziano, \emph{{Classical
  gravitational observables from the Eikonal operator}},
  \href{https://doi.org/10.1016/j.physletb.2023.138049}{\emph{Phys. Lett. B}
  {\bfseries 843} (2023) 138049}
  [\href{https://arxiv.org/abs/2210.12118}{{\ttfamily 2210.12118}}].

\bibitem{DiVecchia:2022owy}
P.~Di~Vecchia, C.~Heissenberg and R.~Russo, \emph{{Angular momentum of
  zero-frequency gravitons}},
  \href{https://doi.org/10.1007/JHEP08(2022)172}{\emph{JHEP} {\bfseries 08}
  (2022) 172} [\href{https://arxiv.org/abs/2203.11915}{{\ttfamily
  2203.11915}}].

\bibitem{Bjerrum-Bohr:2018xdl}
N.E.J.~Bjerrum-Bohr, P.H.~Damgaard, G.~Festuccia, L.~Plant\'e and P.~Vanhove,
  \emph{{General Relativity from Scattering Amplitudes}},
  \href{https://doi.org/10.1103/PhysRevLett.121.171601}{\emph{Phys. Rev. Lett.}
  {\bfseries 121} (2018) 171601}
  [\href{https://arxiv.org/abs/1806.04920}{{\ttfamily 1806.04920}}].

\bibitem{Cristofoli:2019neg}
A.~Cristofoli, N.E.J.~Bjerrum-Bohr, P.H.~Damgaard and P.~Vanhove,
  \emph{{Post-Minkowskian Hamiltonians in general relativity}},
  \href{https://doi.org/10.1103/PhysRevD.100.084040}{\emph{Phys. Rev. D}
  {\bfseries 100} (2019) 084040}
  [\href{https://arxiv.org/abs/1906.01579}{{\ttfamily 1906.01579}}].

\bibitem{Bjerrum-Bohr:2019kec}
N.E.J.~Bjerrum-Bohr, A.~Cristofoli and P.H.~Damgaard, \emph{{Post-Minkowskian
  Scattering Angle in Einstein Gravity}},
  \href{https://doi.org/10.1007/JHEP08(2020)038}{\emph{JHEP} {\bfseries 08}
  (2020) 038} [\href{https://arxiv.org/abs/1910.09366}{{\ttfamily
  1910.09366}}].

\bibitem{Bjerrum-Bohr:2021vuf}
N.E.J.~Bjerrum-Bohr, P.H.~Damgaard, L.~Plant\'e and P.~Vanhove,
  \emph{{Classical gravity from loop amplitudes}},
  \href{https://doi.org/10.1103/PhysRevD.104.026009}{\emph{Phys. Rev. D}
  {\bfseries 104} (2021) 026009}
  [\href{https://arxiv.org/abs/2104.04510}{{\ttfamily 2104.04510}}].

\bibitem{Bjerrum-Bohr:2021din}
N.E.J.~Bjerrum-Bohr, P.H.~Damgaard, L.~Plant\'e and P.~Vanhove, \emph{{The
  amplitude for classical gravitational scattering at third Post-Minkowskian
  order}}, \href{https://doi.org/10.1007/JHEP08(2021)172}{\emph{JHEP}
  {\bfseries 08} (2021) 172}
  [\href{https://arxiv.org/abs/2105.05218}{{\ttfamily 2105.05218}}].

\bibitem{Bjerrum-Bohr:2021wwt}
N.E.J.~Bjerrum-Bohr, L.~Plant\'e and P.~Vanhove, \emph{{Post-Minkowskian radial
  action from soft limits and velocity cuts}},
  \href{https://doi.org/10.1007/JHEP03(2022)071}{\emph{JHEP} {\bfseries 03}
  (2022) 071} [\href{https://arxiv.org/abs/2111.02976}{{\ttfamily
  2111.02976}}].

\bibitem{Menezes:2022tcs}
G.~Menezes and M.~Sergola, \emph{{NLO deflections for spinning particles and
  Kerr black holes}},
  \href{https://doi.org/10.1007/JHEP10(2022)105}{\emph{JHEP} {\bfseries 10}
  (2022) 105} [\href{https://arxiv.org/abs/2205.11701}{{\ttfamily
  2205.11701}}].

\bibitem{Kalin:2019rwq}
G.~K\"alin and R.A.~Porto, \emph{{From Boundary Data to Bound States}},
  \href{https://doi.org/10.1007/JHEP01(2020)072}{\emph{JHEP} {\bfseries 01}
  (2020) 072} [\href{https://arxiv.org/abs/1910.03008}{{\ttfamily
  1910.03008}}].

\bibitem{Cho:2021arx}
G.~Cho, G.~K\"alin and R.A.~Porto, \emph{{From boundary data to bound states.
  Part III. Radiative effects}},
  \href{https://doi.org/10.1007/JHEP04(2022)154}{\emph{JHEP} {\bfseries 04}
  (2022) 154} [\href{https://arxiv.org/abs/2112.03976}{{\ttfamily
  2112.03976}}].

\bibitem{Kalin:2019inp}
G.~K\"alin and R.A.~Porto, \emph{{From boundary data to bound states. Part II.
  Scattering angle to dynamical invariants (with twist)}},
  \href{https://doi.org/10.1007/JHEP02(2020)120}{\emph{JHEP} {\bfseries 02}
  (2020) 120} [\href{https://arxiv.org/abs/1911.09130}{{\ttfamily
  1911.09130}}].

\bibitem{Kalin:2020fhe}
G.~K\"alin, Z.~Liu and R.A.~Porto, \emph{{Conservative Dynamics of Binary
  Systems to Third Post-Minkowskian Order from the Effective Field Theory
  Approach}}, \href{https://doi.org/10.1103/PhysRevLett.125.261103}{\emph{Phys.
  Rev. Lett.} {\bfseries 125} (2020) 261103}
  [\href{https://arxiv.org/abs/2007.04977}{{\ttfamily 2007.04977}}].

\bibitem{Kalin:2020lmz}
G.~K\"alin, Z.~Liu and R.A.~Porto, \emph{{Conservative Tidal Effects in Compact
  Binary Systems to Next-to-Leading Post-Minkowskian Order}},
  \href{https://doi.org/10.1103/PhysRevD.102.124025}{\emph{Phys. Rev. D}
  {\bfseries 102} (2020) 124025}
  [\href{https://arxiv.org/abs/2008.06047}{{\ttfamily 2008.06047}}].

\bibitem{Dlapa:2021npj}
C.~Dlapa, G.~K\"alin, Z.~Liu and R.A.~Porto, \emph{{Dynamics of binary systems
  to fourth Post-Minkowskian order from the effective field theory approach}},
  \href{https://doi.org/10.1016/j.physletb.2022.137203}{\emph{Phys. Lett. B}
  {\bfseries 831} (2022) 137203}
  [\href{https://arxiv.org/abs/2106.08276}{{\ttfamily 2106.08276}}].

\bibitem{Dlapa:2021vgp}
C.~Dlapa, G.~K\"alin, Z.~Liu and R.A.~Porto, \emph{{Conservative Dynamics of
  Binary Systems at Fourth Post-Minkowskian Order in the Large-Eccentricity
  Expansion}},
  \href{https://doi.org/10.1103/PhysRevLett.128.161104}{\emph{Phys. Rev. Lett.}
  {\bfseries 128} (2022) 161104}
  [\href{https://arxiv.org/abs/2112.11296}{{\ttfamily 2112.11296}}].

\bibitem{Dlapa:2022lmu}
C.~Dlapa, G.~K\"alin, Z.~Liu, J.~Neef and R.A.~Porto, \emph{{Radiation Reaction
  and Gravitational Waves at Fourth Post-Minkowskian Order}},
  \href{https://doi.org/10.1103/PhysRevLett.130.101401}{\emph{Phys. Rev. Lett.}
  {\bfseries 130} (2023) 101401}
  [\href{https://arxiv.org/abs/2210.05541}{{\ttfamily 2210.05541}}].

\bibitem{Dlapa:2023hsl}
C.~Dlapa, G.~K\"alin, Z.~Liu and R.A.~Porto, \emph{{Bootstrapping the
  relativistic two-body problem}},
  \href{https://doi.org/10.1007/JHEP08(2023)109}{\emph{JHEP} {\bfseries 08}
  (2023) 109} [\href{https://arxiv.org/abs/2304.01275}{{\ttfamily
  2304.01275}}].

\bibitem{Kalin:2020mvi}
G.~K\"alin and R.A.~Porto, \emph{{Post-Minkowskian Effective Field Theory for
  Conservative Binary Dynamics}},
  \href{https://doi.org/10.1007/JHEP11(2020)106}{\emph{JHEP} {\bfseries 11}
  (2020) 106} [\href{https://arxiv.org/abs/2006.01184}{{\ttfamily
  2006.01184}}].

\bibitem{Liu:2021zxr}
Z.~Liu, R.A.~Porto and Z.~Yang, \emph{{Spin Effects in the Effective Field
  Theory Approach to Post-Minkowskian Conservative Dynamics}},
  \href{https://doi.org/10.1007/JHEP06(2021)012}{\emph{JHEP} {\bfseries 06}
  (2021) 012} [\href{https://arxiv.org/abs/2102.10059}{{\ttfamily
  2102.10059}}].

\bibitem{Kalin:2022hph}
G.~K\"alin, J.~Neef and R.A.~Porto, \emph{{Radiation-reaction in the Effective
  Field Theory approach to Post-Minkowskian dynamics}},
  \href{https://doi.org/10.1007/JHEP01(2023)140}{\emph{JHEP} {\bfseries 01}
  (2023) 140} [\href{https://arxiv.org/abs/2207.00580}{{\ttfamily
  2207.00580}}].

\bibitem{Kosower:2018adc}
D.A.~Kosower, B.~Maybee and D.~O'Connell, \emph{{Amplitudes, Observables, and
  Classical Scattering}},
  \href{https://doi.org/10.1007/JHEP02(2019)137}{\emph{JHEP} {\bfseries 02}
  (2019) 137} [\href{https://arxiv.org/abs/1811.10950}{{\ttfamily
  1811.10950}}].

\bibitem{Cristofoli:2021vyo}
A.~Cristofoli, R.~Gonzo, D.A.~Kosower and D.~O'Connell, \emph{{Waveforms from
  amplitudes}}, \href{https://doi.org/10.1103/PhysRevD.106.056007}{\emph{Phys.
  Rev. D} {\bfseries 106} (2022) 056007}
  [\href{https://arxiv.org/abs/2107.10193}{{\ttfamily 2107.10193}}].

\bibitem{Maybee:2019jus}
B.~Maybee, D.~O'Connell and J.~Vines, \emph{{Observables and amplitudes for
  spinning particles and black holes}},
  \href{https://doi.org/10.1007/JHEP12(2019)156}{\emph{JHEP} {\bfseries 12}
  (2019) 156} [\href{https://arxiv.org/abs/1906.09260}{{\ttfamily
  1906.09260}}].

\bibitem{Aoude:2021oqj}
R.~Aoude and A.~Ochirov, \emph{{Classical observables from coherent-spin
  amplitudes}}, \href{https://doi.org/10.1007/JHEP10(2021)008}{\emph{JHEP}
  {\bfseries 10} (2021) 008}
  [\href{https://arxiv.org/abs/2108.01649}{{\ttfamily 2108.01649}}].

\bibitem{Adamo:2022rmp}
T.~Adamo, A.~Cristofoli and A.~Ilderton, \emph{{Classical physics from
  amplitudes on curved backgrounds}},
  \href{https://doi.org/10.1007/JHEP08(2022)281}{\emph{JHEP} {\bfseries 08}
  (2022) 281} [\href{https://arxiv.org/abs/2203.13785}{{\ttfamily
  2203.13785}}].

\bibitem{Sen:2017nim}
A.~Sen, \emph{{Subleading Soft Graviton Theorem for Loop Amplitudes}},
  \href{https://doi.org/10.1007/JHEP11(2017)123}{\emph{JHEP} {\bfseries 11}
  (2017) 123} [\href{https://arxiv.org/abs/1703.00024}{{\ttfamily
  1703.00024}}].

\bibitem{Laddha:2017ygw}
A.~Laddha and A.~Sen, \emph{{Sub-subleading Soft Graviton Theorem in Generic
  Theories of Quantum Gravity}},
  \href{https://doi.org/10.1007/JHEP10(2017)065}{\emph{JHEP} {\bfseries 10}
  (2017) 065} [\href{https://arxiv.org/abs/1706.00759}{{\ttfamily
  1706.00759}}].

\bibitem{Chakrabarti:2017ltl}
S.~Chakrabarti, S.P.~Kashyap, B.~Sahoo, A.~Sen and M.~Verma, \emph{{Subleading
  Soft Theorem for Multiple Soft Gravitons}},
  \href{https://doi.org/10.1007/JHEP12(2017)150}{\emph{JHEP} {\bfseries 12}
  (2017) 150} [\href{https://arxiv.org/abs/1707.06803}{{\ttfamily
  1707.06803}}].

\bibitem{Laddha:2018rle}
A.~Laddha and A.~Sen, \emph{{Gravity Waves from Soft Theorem in General
  Dimensions}}, \href{https://doi.org/10.1007/JHEP09(2018)105}{\emph{JHEP}
  {\bfseries 09} (2018) 105}
  [\href{https://arxiv.org/abs/1801.07719}{{\ttfamily 1801.07719}}].

\bibitem{Laddha:2018myi}
A.~Laddha and A.~Sen, \emph{{Logarithmic Terms in the Soft Expansion in Four
  Dimensions}}, \href{https://doi.org/10.1007/JHEP10(2018)056}{\emph{JHEP}
  {\bfseries 10} (2018) 056}
  [\href{https://arxiv.org/abs/1804.09193}{{\ttfamily 1804.09193}}].

\bibitem{Sahoo:2018lxl}
B.~Sahoo and A.~Sen, \emph{{Classical and Quantum Results on Logarithmic Terms
  in the Soft Theorem in Four Dimensions}},
  \href{https://doi.org/10.1007/JHEP02(2019)086}{\emph{JHEP} {\bfseries 02}
  (2019) 086} [\href{https://arxiv.org/abs/1808.03288}{{\ttfamily
  1808.03288}}].

\bibitem{Laddha:2019yaj}
A.~Laddha and A.~Sen, \emph{{Classical proof of the classical soft graviton
  theorem in D\ensuremath{>}4}},
  \href{https://doi.org/10.1103/PhysRevD.101.084011}{\emph{Phys. Rev. D}
  {\bfseries 101} (2020) 084011}
  [\href{https://arxiv.org/abs/1906.08288}{{\ttfamily 1906.08288}}].

\bibitem{Saha:2019tub}
A.P.~Saha, B.~Sahoo and A.~Sen, \emph{{Proof of the classical soft graviton
  theorem in $D$ = 4}},
  \href{https://doi.org/10.1007/JHEP06(2020)153}{\emph{JHEP} {\bfseries 06}
  (2020) 153} [\href{https://arxiv.org/abs/1912.06413}{{\ttfamily
  1912.06413}}].

\bibitem{Ghosh:2021bam}
D.~Ghosh and B.~Sahoo, \emph{{Spin-dependent gravitational tail memory in
  $D=4$}}, \href{https://doi.org/10.1103/PhysRevD.105.025024}{\emph{Phys. Rev.
  D} {\bfseries 105} (2022) 025024}
  [\href{https://arxiv.org/abs/2106.10741}{{\ttfamily 2106.10741}}].

\bibitem{Sahoo:2020ryf}
B.~Sahoo, \emph{{Classical Sub-subleading Soft Photon and Soft Graviton
  Theorems in Four Spacetime Dimensions}},
  \href{https://doi.org/10.1007/JHEP12(2020)070}{\emph{JHEP} {\bfseries 12}
  (2020) 070} [\href{https://arxiv.org/abs/2008.04376}{{\ttfamily
  2008.04376}}].

\bibitem{nj}
E.T.~Newman and A.I.~Janis, \emph{{Note on the Kerr Spinning‐Particle
  Metric}}, {\emph{Journal of Mathematical Physics} {\bfseries 6} (2004) 915}.

\bibitem{Erbin:2016lzq}
H.~Erbin, \emph{{Janis-Newman algorithm: generating rotating and NUT charged
  black holes}}, \href{https://doi.org/10.3390/universe3010019}{\emph{Universe}
  {\bfseries 3} (2017) 19} [\href{https://arxiv.org/abs/1701.00037}{{\ttfamily
  1701.00037}}].

\bibitem{Lynden-Bell:2002dvr}
D.~Lynden-Bell, \emph{{A magic electromagnetic field}},
  \href{https://arxiv.org/abs/astro-ph/0207064}{{\ttfamily astro-ph/0207064}}.

\bibitem{Alessio:2023kgf}
F.~Alessio, \emph{{Kerr binary dynamics from minimal coupling and double
  copy}},  \href{https://arxiv.org/abs/2303.12784}{{\ttfamily 2303.12784}}.

\bibitem{Bern:2023ity}
Z.~Bern, D.~Kosmopoulos, A.~Luna, R.~Roiban, T.~Scheopner, F.~Teng et~al.,
  \emph{{Quantum Field Theory, Worldline Theory, and Spin Magnitude Change in
  Orbital Evolution}},  \href{https://arxiv.org/abs/2308.14176}{{\ttfamily
  2308.14176}}.

\bibitem{Bautista:2021llr}
Y.F.~Bautista and A.~Laddha, \emph{{Soft constraints on KMOC formalism}},
  \href{https://doi.org/10.1007/JHEP12(2022)018}{\emph{JHEP} {\bfseries 12}
  (2022) 018} [\href{https://arxiv.org/abs/2111.11642}{{\ttfamily
  2111.11642}}].

\bibitem{Elkhidir:2023dco}
A.~Elkhidir, D.~O'Connell, M.~Sergola and I.A.~Vazquez-Holm, \emph{{Radiation
  and Reaction at One Loop}},
  \href{https://arxiv.org/abs/2303.06211}{{\ttfamily 2303.06211}}.

\bibitem{delaCruz:2020bbn}
L.~de~la Cruz, B.~Maybee, D.~O'Connell and A.~Ross, \emph{{Classical Yang-Mills
  observables from amplitudes}},
  \href{https://doi.org/10.1007/JHEP12(2020)076}{\emph{JHEP} {\bfseries 12}
  (2020) 076} [\href{https://arxiv.org/abs/2009.03842}{{\ttfamily
  2009.03842}}].

\bibitem{Adamo:2021rfq}
T.~Adamo, A.~Cristofoli and P.~Tourkine, \emph{{Eikonal amplitudes from curved
  backgrounds}},
  \href{https://doi.org/10.21468/SciPostPhys.13.2.032}{\emph{SciPost Phys.}
  {\bfseries 13} (2022) 032}
  [\href{https://arxiv.org/abs/2112.09113}{{\ttfamily 2112.09113}}].

\bibitem{Guevara:2020xjx}
A.~Guevara, B.~Maybee, A.~Ochirov, D.~O'connell and J.~Vines, \emph{{A
  worldsheet for Kerr}},
  \href{https://doi.org/10.1007/JHEP03(2021)201}{\emph{JHEP} {\bfseries 03}
  (2021) 201} [\href{https://arxiv.org/abs/2012.11570}{{\ttfamily
  2012.11570}}].

\bibitem{Chung:2019yfs}
M.-Z.~Chung, Y.-T.~Huang and J.-W.~Kim, \emph{{Kerr-Newman stress-tensor from
  minimal coupling}},
  \href{https://doi.org/10.1007/JHEP12(2020)103}{\emph{JHEP} {\bfseries 12}
  (2020) 103} [\href{https://arxiv.org/abs/1911.12775}{{\ttfamily
  1911.12775}}].

\bibitem{A:2022wsk}
M.~A. and D.~Ghosh, \emph{{Classical spinning soft factors from gauge theory
  amplitudes}},  \href{https://arxiv.org/abs/2210.07561}{{\ttfamily
  2210.07561}}.

\bibitem{Vines:2017hyw}
J.~Vines, \emph{{Scattering of two spinning black holes in post-Minkowskian
  gravity, to all orders in spin, and effective-one-body mappings}},
  \href{https://doi.org/10.1088/1361-6382/aaa3a8}{\emph{Class. Quant. Grav.}
  {\bfseries 35} (2018) 084002}
  [\href{https://arxiv.org/abs/1709.06016}{{\ttfamily 1709.06016}}].

\bibitem{Arkani-Hamed:2017jhn}
N.~Arkani-Hamed, T.-C.~Huang and Y.-t.~Huang, \emph{{Scattering amplitudes for
  all masses and spins}},
  \href{https://doi.org/10.1007/JHEP11(2021)070}{\emph{JHEP} {\bfseries 11}
  (2021) 070} [\href{https://arxiv.org/abs/1709.04891}{{\ttfamily
  1709.04891}}].

\bibitem{Chung:2018kqs}
M.-Z.~Chung, Y.-T.~Huang, J.-W.~Kim and S.~Lee, \emph{{The simplest massive
  S-matrix: from minimal coupling to Black Holes}},
  \href{https://doi.org/10.1007/JHEP04(2019)156}{\emph{JHEP} {\bfseries 04}
  (2019) 156} [\href{https://arxiv.org/abs/1812.08752}{{\ttfamily
  1812.08752}}].

\bibitem{Guevara:2018wpp}
A.~Guevara, A.~Ochirov and J.~Vines, \emph{{Scattering of Spinning Black Holes
  from Exponentiated Soft Factors}},
  \href{https://doi.org/10.1007/JHEP09(2019)056}{\emph{JHEP} {\bfseries 09}
  (2019) 056} [\href{https://arxiv.org/abs/1812.06895}{{\ttfamily
  1812.06895}}].

\bibitem{Guevara:2017csg}
A.~Guevara, \emph{{Holomorphic Classical Limit for Spin Effects in
  Gravitational and Electromagnetic Scattering}},
  \href{https://doi.org/10.1007/JHEP04(2019)033}{\emph{JHEP} {\bfseries 04}
  (2019) 033} [\href{https://arxiv.org/abs/1706.02314}{{\ttfamily
  1706.02314}}].

\bibitem{Aoude:2020onz}
R.~Aoude, K.~Haddad and A.~Helset, \emph{{On-shell heavy particle effective
  theories}}, \href{https://doi.org/10.1007/JHEP05(2020)051}{\emph{JHEP}
  {\bfseries 05} (2020) 051}
  [\href{https://arxiv.org/abs/2001.09164}{{\ttfamily 2001.09164}}].

\bibitem{Haddad:2020tvs}
K.~Haddad and A.~Helset, \emph{{The double copy for heavy particles}},
  \href{https://doi.org/10.1103/PhysRevLett.125.181603}{\emph{Phys. Rev. Lett.}
  {\bfseries 125} (2020) 181603}
  [\href{https://arxiv.org/abs/2005.13897}{{\ttfamily 2005.13897}}].

\bibitem{Chiodaroli:2021eug}
M.~Chiodaroli, H.~Johansson and P.~Pichini, \emph{{Compton black-hole
  scattering for s \ensuremath{\leq} 5/2}},
  \href{https://doi.org/10.1007/JHEP02(2022)156}{\emph{JHEP} {\bfseries 02}
  (2022) 156} [\href{https://arxiv.org/abs/2107.14779}{{\ttfamily
  2107.14779}}].

\bibitem{Aoude:2022trd}
R.~Aoude, K.~Haddad and A.~Helset, \emph{{Searching for Kerr in the 2PM
  amplitude}}, \href{https://doi.org/10.1007/JHEP07(2022)072}{\emph{JHEP}
  {\bfseries 07} (2022) 072}
  [\href{https://arxiv.org/abs/2203.06197}{{\ttfamily 2203.06197}}].

\bibitem{Aoude:2022thd}
R.~Aoude, K.~Haddad and A.~Helset, \emph{{Classical Gravitational
  Spinning-Spinless Scattering at O(G2S\ensuremath{\infty})}},
  \href{https://doi.org/10.1103/PhysRevLett.129.141102}{\emph{Phys. Rev. Lett.}
  {\bfseries 129} (2022) 141102}
  [\href{https://arxiv.org/abs/2205.02809}{{\ttfamily 2205.02809}}].

\bibitem{Chen:2022clh}
W.-M.~Chen, M.-Z.~Chung, Y.-t.~Huang and J.-W.~Kim, \emph{{Gravitational
  Faraday effect from on-shell amplitudes}},
  \href{https://doi.org/10.1007/JHEP12(2022)058}{\emph{JHEP} {\bfseries 12}
  (2022) 058} [\href{https://arxiv.org/abs/2205.07305}{{\ttfamily
  2205.07305}}].

\bibitem{Cangemi:2022abk}
L.~Cangemi and P.~Pichini, \emph{{Classical limit of higher-spin string
  amplitudes}}, \href{https://doi.org/10.1007/JHEP06(2023)167}{\emph{JHEP}
  {\bfseries 06} (2023) 167}
  [\href{https://arxiv.org/abs/2207.03947}{{\ttfamily 2207.03947}}].

\bibitem{Saketh:2022wap}
M.V.S.~Saketh and J.~Vines, \emph{{Scattering of gravitational waves off
  spinning compact objects with an effective worldline theory}},
  \href{https://doi.org/10.1103/PhysRevD.106.124026}{\emph{Phys. Rev. D}
  {\bfseries 106} (2022) 124026}
  [\href{https://arxiv.org/abs/2208.03170}{{\ttfamily 2208.03170}}].

\bibitem{Bjerrum-Bohr:2023jau}
N.E.J.~Bjerrum-Bohr, G.~Chen and M.~Skowronek, \emph{{Classical spin
  gravitational Compton scattering}},
  \href{https://doi.org/10.1007/JHEP06(2023)170}{\emph{JHEP} {\bfseries 06}
  (2023) 170} [\href{https://arxiv.org/abs/2302.00498}{{\ttfamily
  2302.00498}}].

\bibitem{Bjerrum-Bohr:2023iey}
N.E.J.~Bjerrum-Bohr, G.~Chen and M.~Skowronek, \emph{{Covariant Compton
  Amplitudes in Gravity with Classical Spin}},
  \href{https://arxiv.org/abs/2309.11249}{{\ttfamily 2309.11249}}.

\bibitem{Haddad:2023ylx}
K.~Haddad, \emph{{Recursion in the classical limit and the neutron-star Compton
  amplitude}}, \href{https://doi.org/10.1007/JHEP05(2023)177}{\emph{JHEP}
  {\bfseries 05} (2023) 177}
  [\href{https://arxiv.org/abs/2303.02624}{{\ttfamily 2303.02624}}].

\bibitem{Brandhuber:2023hhy}
A.~Brandhuber, G.R.~Brown, G.~Chen, S.~De~Angelis, J.~Gowdy and G.~Travaglini,
  \emph{{One-loop gravitational bremsstrahlung and waveforms from a heavy-mass
  effective field theory}},
  \href{https://doi.org/10.1007/JHEP06(2023)048}{\emph{JHEP} {\bfseries 06}
  (2023) 048} [\href{https://arxiv.org/abs/2303.06111}{{\ttfamily
  2303.06111}}].

\bibitem{Aoude:2023vdk}
R.~Aoude, K.~Haddad and A.~Helset, \emph{{Classical gravitational scattering
  amplitude at O(G2S1\ensuremath{\infty}S2\ensuremath{\infty})}},
  \href{https://doi.org/10.1103/PhysRevD.108.024050}{\emph{Phys. Rev. D}
  {\bfseries 108} (2023) 024050}
  [\href{https://arxiv.org/abs/2304.13740}{{\ttfamily 2304.13740}}].

\bibitem{Brandhuber:2023hhl}
A.~Brandhuber, G.R.~Brown, G.~Chen, J.~Gowdy and G.~Travaglini, \emph{{Resummed
  spinning waveforms from five-point amplitudes}},
  \href{https://doi.org/10.1007/JHEP02(2024)026}{\emph{JHEP} {\bfseries 02}
  (2024) 026} [\href{https://arxiv.org/abs/2310.04405}{{\ttfamily
  2310.04405}}].

\bibitem{Scheopner:2023rzp}
T.~Scheopner and J.~Vines, \emph{{Dynamical Implications of the Kerr Multipole
  Moments for Spinning Black Holes}},
  \href{https://arxiv.org/abs/2311.18421}{{\ttfamily 2311.18421}}.

\bibitem{Bautista:2021wfy}
Y.F.~Bautista, A.~Guevara, C.~Kavanagh and J.~Vines, \emph{{Scattering in black
  hole backgrounds and higher-spin amplitudes. Part I}},
  \href{https://doi.org/10.1007/JHEP03(2023)136}{\emph{JHEP} {\bfseries 03}
  (2023) 136} [\href{https://arxiv.org/abs/2107.10179}{{\ttfamily
  2107.10179}}].

\bibitem{Bautista:2022wjf}
Y.F.~Bautista, A.~Guevara, C.~Kavanagh and J.~Vines, \emph{{Scattering in black
  hole backgrounds and higher-spin amplitudes. Part II}},
  \href{https://doi.org/10.1007/JHEP05(2023)211}{\emph{JHEP} {\bfseries 05}
  (2023) 211} [\href{https://arxiv.org/abs/2212.07965}{{\ttfamily
  2212.07965}}].

\bibitem{Bautista:2023szu}
Y.F.~Bautista, \emph{{Dynamics for super-extremal Kerr binary systems at
  O(G2)}}, \href{https://doi.org/10.1103/PhysRevD.108.084036}{\emph{Phys. Rev.
  D} {\bfseries 108} (2023) 084036}
  [\href{https://arxiv.org/abs/2304.04287}{{\ttfamily 2304.04287}}].

\bibitem{Bautista:2023sdf}
Y.F.~Bautista, G.~Bonelli, C.~Iossa, A.~Tanzini and Z.~Zhou, \emph{{Black Hole
  Perturbation Theory Meets CFT$_2$: Kerr Compton Amplitudes from
  Nekrasov-Shatashvili Functions}},
  \href{https://arxiv.org/abs/2312.05965}{{\ttfamily 2312.05965}}.

\bibitem{Akhtar:2024rzp}
S.~Akhtar, A.~Manna and A.~Manu, \emph{{Radiative gauge field using
  Exponentiated Spin factors: Electromagnetic Scattering}},
  \href{https://arxiv.org/abs/in progress}{{\ttfamily in progress}}.

\bibitem{Gralla:2021eoi}
S.E.~Gralla and K.~Lobo, \emph{{Electromagnetic scoot}},
  \href{https://doi.org/10.1103/PhysRevD.105.084053}{\emph{Phys. Rev. D}
  {\bfseries 105} (2022) 084053}
  [\href{https://arxiv.org/abs/2112.01729}{{\ttfamily 2112.01729}}].

\bibitem{Luna:2023uwd}
A.~Luna, N.~Moynihan, D.~O'Connell and A.~Ross, \emph{{Observables from the
  Spinning Eikonal}},  \href{https://arxiv.org/abs/2312.09960}{{\ttfamily
  2312.09960}}.

\bibitem{Bhardwaj:2022hip}
R.~Bhardwaj and L.~Lippstreu, \emph{{Angular momentum of the asymptotic
  electromagnetic field in the classical scattering of charged particles}},
  \href{https://arxiv.org/abs/2208.02727}{{\ttfamily 2208.02727}}.

\bibitem{Bern:2019prr}
Z.~Bern, J.J.~Carrasco, M.~Chiodaroli, H.~Johansson and R.~Roiban, \emph{{The
  Duality Between Color and Kinematics and its Applications}},
  \href{https://arxiv.org/abs/1909.01358}{{\ttfamily 1909.01358}}.

\bibitem{DeAngelis:2023lvf}
S.~De~Angelis, R.~Gonzo and P.P.~Novichkov, \emph{{Spinning waveforms from KMOC
  at leading order}},  \href{https://arxiv.org/abs/2309.17429}{{\ttfamily
  2309.17429}}.

\bibitem{Bohnenblust:2023qmy}
L.~Bohnenblust, H.~Ita, M.~Kraus and J.~Schlenk, \emph{{Gravitational
  Bremsstrahlung in Black-Hole Scattering at $\mathcal{O}(G^3)$: Linear-in-Spin
  Effects}},  \href{https://arxiv.org/abs/2312.14859}{{\ttfamily 2312.14859}}.

\bibitem{Aoude:2023dui}
R.~Aoude, K.~Haddad, C.~Heissenberg and A.~Helset, \emph{{Leading-order
  gravitational radiation to all spin orders}},
  \href{https://doi.org/10.1103/PhysRevD.109.036007}{\emph{Phys. Rev. D}
  {\bfseries 109} (2024) 036007}
  [\href{https://arxiv.org/abs/2310.05832}{{\ttfamily 2310.05832}}].

\bibitem{Manu:2020zxl}
A.~Manu, D.~Ghosh, A.~Laddha and P.V.~Athira, \emph{{Soft radiation from
  scattering amplitudes revisited}},
  \href{https://doi.org/10.1007/JHEP05(2021)056}{\emph{JHEP} {\bfseries 05}
  (2021) 056} [\href{https://arxiv.org/abs/2007.02077}{{\ttfamily
  2007.02077}}].

\end{thebibliography}\endgroup

\end{document}